\documentclass[aps,prd,twocolumn,showpacs]{revtex4-1}
\usepackage{amsmath}
\usepackage{graphicx}
\usepackage{subfigure}
\usepackage{epstopdf}
\usepackage{color}
\usepackage{multirow}
\usepackage{setspace}
\usepackage{overpic}
\usepackage{amssymb}
\usepackage[dvipdfmx, bookmarksnumbered, pdfstartview=FitH,colorlinks,urlcolor=blue, citecolor=blue,linkcolor=blue,] {hyperref}
\usepackage{lineno}
\usepackage{bm}
\usepackage{rotating}
\usepackage[utf8]{inputenc}

\usepackage{array}
\newcommand{\PreserveBackslash}[1]{\let\temp=\\#1\let\\=\temp}
\newcolumntype{C}[1]{>{\PreserveBackslash\centering}p{#1}}
\newcolumntype{R}[1]{>{\PreserveBackslash\raggedleft}p{#1}}
\newcolumntype{L}[1]{>{\PreserveBackslash\raggedright}p{#1}}

\newcommand{\wbrksk}   {$\mathcal{B}(D^+\rightarrow K_S^0 K^+)$ }
\newcommand{\wbrkskpi} {$\mathcal{B}(D^+\rightarrow K_S^0 K^+ \pi^0)$ }
\newcommand{\wbrklk}   {$\mathcal{B}(D^+\rightarrow K_L^0 K^+)$ }
\newcommand{\wbrklkpi} {$\mathcal{B}(D^+\rightarrow K_L^0 K^+ \pi^0)$ }

\newcommand{\brksk}    {$(3.02    \pm  0.09  \pm  0.08)\times10^{-3}$}
\newcommand{\brkskpi}  {$(5.07    \pm  0.19  \pm  0.23)\times10^{-3}$}
\newcommand{\brklk}    {$(3.21    \pm  0.11  \pm  0.11)\times10^{-3}$}
\newcommand{\brklkpi}  {$(5.24    \pm  0.22  \pm  0.22)\times10^{-3}$}

\begin{document}

\graphicspath{{figure/}}
\DeclareGraphicsExtensions{.eps,.png,.ps}

\title{\boldmath Measurements of the absolute branching fractions and $CP$ asymmetries {\color{black}{for}} $D^+\rightarrow K_{S,L}^0 K^+(\pi^0)$}

\author{
  \begin{small}
    \begin{center}
      M.~Ablikim$^{1}$, M.~N.~Achasov$^{9,d}$, S. ~Ahmed$^{14}$, M.~Albrecht$^{4}$, A.~Amoroso$^{50A,50C}$, F.~F.~An$^{1}$, Q.~An$^{47,39}$, J.~Z.~Bai$^{1}$, O.~Bakina$^{24}$, R.~Baldini Ferroli$^{20A}$
% ...
      \\
      \vspace{0.2cm}
      (BESIII Collaboration)\\
      \vspace{0.2cm} {\it
        $^{1}$ Institute of High Energy Physics, Beijing 100049, People's Republic of China\\
% ...
$^{4}$ Bochum Ruhr-University, D-44780 Bochum, Germany\\
% ...
$^{9}$ G.I. Budker Institute of Nuclear Physics SB RAS (BINP), Novosibirsk 630090, Russia\\
% ...
$^{14}$ Helmholtz Institute Mainz, Johann-Joachim-Becher-Weg 45, D-55099 Mainz, Germany\\
% ...
$^{20}$ (A)INFN Laboratori Nazionali di Frascati, I-00044, Frascati, Italy; (B)INFN and University of Perugia, I-06100, Perugia, Italy\\
% ...
$^{24}$ Joint Institute for Nuclear Research, 141980 Dubna, Moscow region, Russia\\
% ...
$^{39}$ State Key Laboratory of Particle Detection and Electronics, Beijing 100049, Hefei 230026, People's Republic of China\\
% ...
$^{47}$ University of Science and Technology of China, Hefei 230026, People's Republic of China\\
% ...
$^{50}$ (A)University of Turin, I-10125, Turin, Italy; (B)University of Eastern Piedmont, I-15121, Alessandria, Italy; (C)INFN, I-10125, Turin, Italy\\
% ...
$^{d}$ Also at the Novosibirsk State University, Novosibirsk, 630090, Russia\\
% ...
      }\end{center}
    \vspace{0.4cm}
  \end{small}
}

\affiliation{}

\date{\today}

\begin{spacing}{1.0}

\begin{abstract}
Using  $e^+e^-$ collision data corresponding to an integrated
luminosity of 2.93 fb$^{-1}$ taken at a center-of-mass energy of 3.773 GeV with the BESIII detector, we determine the absolute branching fractions \wbrksk = \brksk, \wbrkskpi = \brkskpi, \wbrklk = \brklk, and \wbrklkpi = \brklkpi, where the first and second uncertainties are statistical and systematic, respectively. The branching fraction \wbrksk is consistent with the world average value and the other three branching fractions are measured for the first time. We also measure the $CP$ asymmetries for the four decays and do not find a significant deviation from zero.
\end{abstract}


\maketitle

\section{Introduction}

Experimental studies of hadronic decays of charm mesons shed light on the interplay between the strong and weak forces. In the standard model (SM), the singly Cabibbo-suppressed (SCS) $D^\pm$ meson hadronic decays are predicted to exhibit $CP$ asymmetries of the order of $10^{-3}$~\cite{cpscs1}. Direct $CP$ violation in SCS $D^\pm$ meson decays can arise from the interference between tree-level and penguin decay processes \cite{SU3}. However, the doubly Cabibbo-suppressed and Cabibbo-favored $D^\pm$ meson decays are expected to be $CP$ invariant because they are dominated by a single weak amplitude. Consequently, any observation of $CP$ asymmetry greater than $\mathcal{O}$($10^{-3}$) in the SCS $D^\pm$ meson hadronic decays would be evidence for new physics beyond the SM \cite{NonLep}. In theory, the branching fractions of two-body hadronic decays of $D$ mesons can be calculated within SU(3) flavor symmetry~\cite{DtoPP}. An improved measurement of the branching fraction of the SCS decay $D^+ \rightarrow \bar K^0 K^+$ will help to test the theoretical calculations and benefit the understanding of the violation of SU(3) flavor symmetry in $D$ meson decays \cite{DtoPP}. In this paper, we present measurements of the absolute branching fractions and the direct $CP$ asymmetries of the SCS decays of $D^+ \rightarrow K^0_SK^+$,  $K^0_SK^+\pi^0$, $K^0_LK^+$ and $K^0_LK^+\pi^0$.

In this analysis, we employ the ``double-tag" (DT) technique, which was first developed by the MARK-III Collaboration~\cite{DT1,DT2}, to measure the absolute branching fractions. First, we select ``single-tag" (ST) events in which either a $D$ or $\bar D$ meson is fully reconstructed in one of several specific hadronic decays. Then we look for the $D$ meson decays of interest in the presence of the ST $\bar D$ events; the so called the DT events in which both the $D$ and $\bar D$ mesons are fully reconstructed. The ST and DT yields ($N_{\rm ST}$ and $N_{\rm DT}$) can be described by

\begin{equation}
\begin{aligned}
&N_{\rm ST} = 2N_{D^+D^-} \mathcal{B}_{\rm tag} \epsilon_{\rm ST},\\
&N_{\rm DT} = 2N_{D^+D^-} \mathcal{B}_{\rm tag} \mathcal{B}_{\rm sig} \epsilon_{\rm DT},
\end{aligned}
\end{equation}

\noindent where $N_{D^+D^-}$ is the total number of $D^+D^-$ pairs produced in data, $\epsilon_{\rm ST}$ and $\epsilon_{\rm DT}$ are the efficiencies of reconstructing the ST and DT candidate events, and $\mathcal{B}_{\rm tag}$ and $\mathcal{B}_{\rm sig}$ are the branching fractions for the tag mode and the signal mode, respectively. The absolute branching fraction for the signal decay can be determined by

\begin{equation}\label{BF}
\mathcal{B}_{\rm sig} = \frac {N_{\rm DT}/\epsilon_{\rm DT}} {N_{\rm ST}/\epsilon_{\rm ST}} = \frac {N_{\rm DT}/\epsilon} {N_{\rm ST}} ,
\end{equation}

\noindent where $\epsilon = \epsilon_{\rm DT}/\epsilon_{\rm ST}$  is the efficiency of finding a signal candidate in the presence of a ST $\bar D$, which can be obtained from MC simulations.

With the measured absolute branching fractions of $D^+$ and $D^-$ meson decays ($\mathcal{B}_{\rm sig}^+$ and $\mathcal{B}_{\rm sig}^-$), the $CP$ asymmetry for the decay of interest can be determined by

\begin{equation}\label{acp}
\mathcal{A}_{CP} = \frac {\mathcal{B}_{\rm sig}^+ - \mathcal{B}_{\rm sig}^-}   {\mathcal{B}_{\rm sig}^+ + \mathcal{B}_{\rm sig}^-}.
\end{equation}

\section{THE BESIII DETECTOR AND DATA SAMPLE}

The analysis presented in this paper is based on a data sample with an integrated luminosity of 2.93 fb$^{-1}$ \cite{lum} collected with the BESIII detector \cite{detector} at the center-of-mass (c.m.) energy of $\sqrt s = 3.773$~GeV. The BESIII detector is a general-purpose detector at the BEPCII \cite{BEPCII} with double storage rings. The detector has a geometrical acceptance of 93$\%$ of the full solid angle. We briefly describe the components of BESIII from the interaction point (IP) outward. A small-cell multi-layer drift chamber (MDC), using a helium-based gas to measure momenta and specific ionization of charged particles, is surrounded by a time-of-flight (TOF) system based on plastic scintillators which determines the time of flight of charged particles. A CsI(Tl) electromagnetic calorimeter (EMC) detects electromagnetic showers. These components are all situated inside a superconducting solenoid magnet, which provides a 1.0 T magnetic field parallel to the beam direction. Finally, a multilayer resistive plate counter system installed in the iron flux return yoke of the magnet is used to track muons. The momentum resolution for charged tracks in the MDC is 0.5$\%$ for a transverse momentum of 1 GeV$/c$. The specific energy loss ($dE/dx$) measured in the MDC has a resolution better than $6\%$. The TOF can measure the flight time of charged particles with a time resolution of 80 ps in the barrel and 110 ps in the end caps. The energy resolution for the EMC is $2.5\%$ in the barrel and $5.0\%$ in the end caps for photons and electrons with an energy of 1 GeV. The position resolution of the EMC is 6 mm in the barrel and 9 mm in the end caps. More details on the features and capabilities of BESIII can be found elsewhere \cite{detector}.

A {\sc geant4}-based \cite{geant4} Monte Carlo (MC) simulation software package, which includes the geometric description of the detector and its response, is used to determine the detector efficiency and to estimate potential backgrounds. An inclusive MC sample, which includes the $D^0 \bar D^0$, $D^+ D^-$, and non-$D \bar D$ decays of $\psi(3770)$, the initial state radiation (ISR) production of $\psi(3686)$ and $J/\psi$, the $q \bar q$ ($q = u, d, s$) continuum process, Bhabha scattering events, and di-muon and di-tau events, is produced at $\sqrt s = 3.773$~GeV. The {\sc kkmc} \cite{KKMC} package, which incorporates the beam energy spread and the ISR effects (radiative corrections up to next to leading order), is used to generate the $\psi(3770)$ meson. Final state radiation of charged tracks is simulated with the {\sc photos} package \cite{FSR}. $\psi(3770) \rightarrow D \bar D$ events are generated using {\sc evtgen} \cite{EVENT1, EVENT2}, and each $D$ meson is allowed to decay according to the branching fractions in the Particle Data Group (PDG) \cite{PDG}. This sample is referred as the ``generic" MC sample. Another MC sample of $\psi(3770) \rightarrow D \bar D$ events, in which one $D$ meson decays to the signal mode and the other one decays to any of the ST modes, is referred as the ``signal" MC sample. In both the generic and signal MC samples, the two-body decays $D^+\to K^0_{S,L}K^+$ are generated with a phase space model, while the three-body decays $D^+\to K^0_{S,L}K^+\pi^0$ are generated as a mixture of known intermediate decays with fractions taken from the Dalitz plot analysis of their charge conjugated decay $D^+ \rightarrow K^+K^-\pi^+$ \cite{kkpi}.

\section{DATA ANALYSIS}

The ST $D^{\mp}$ mesons are reconstructed using six hadronic final states: $K^{\pm}\pi^{\mp}\pi^{\mp}$, $K^{\pm}\pi^{\mp}\pi^{\mp}\pi^0$, $K_S^0\pi^{\mp}$, $K_S^0\pi^{\mp}\pi^0$, $K_S^0\pi^{\pm}\pi^{\mp}\pi^{\mp}$ and $K^{\mp}K^{\pm}\pi^{\mp}$, where $ K_S^0$ is reconstructed by its $\pi^+\pi^-$ decay mode and $\pi^0$ with the $\gamma\gamma$ final state. The event selection criteria are described below.

Charged tracks are reconstructed within the MDC coverage $|\cos \theta | < 0.93$, where $\theta$ is the polar angle with respect to the positron beam direction. Tracks (except for those from $K_S^0$ decays) are required to have a point of closest approach to the IP satisfying $|V_z| < 10$ cm in the beam direction and $|V_r| < 1$~cm in the plane perpendicular to the beam direction. Particle identification (PID) is performed by combining the information of $dE/dx$ in the MDC and the flight time obtained from the TOF.  For a charged $\pi(K)$ candidate, the probability of the $\pi (K)$ hypothesis is required to be larger than that of the $K (\pi)$ hypothesis.

The $K_S^0$ candidates are reconstructed from combinations of two tracks with opposite charges which satisfy $|V_z| < 20$~cm, but without requirement on $|V_r|$. The two charged tracks are assumed to be $\pi^+\pi^-$ without PID and are constrained to originate from a common decay vertex. The $\pi^+\pi^-$ invariant masses $M_{\pi^+\pi^-}$ are required to satisfy $|M_{\pi^+\pi^-} - M_{K_S^0}| < 12$~MeV/$c^2$, where $M_{K_S^0}$ is the nominal $K_S^0$ mass \cite{PDG}. Finally, the $K_S^0$ candidates are required to have a decay length significance $L / \sigma_L$ of more than two standard deviations, as obtained from the vertex fit.

Photon candidates are selected from isolated showers in the EMC with minimum energy larger than 25 MeV in the barrel region $(|\cos \theta | < 0.80)$ or 50 MeV in the end-cap region $(0.86 <|\cos \theta | < 0.92)$. The shower timing is required to be no later than 700~ns after the event start time to suppress electronic noise and energy deposits unrelated to the event.

The $\pi^0$ candidates are reconstructed from pairs of photon candidates with invariant mass within $0.110 < M_{\gamma\gamma} < 0.155$ GeV/$c^2$. The $\gamma \gamma$ invariant mass is then constrained to the nominal $\pi^0$ mass \cite{PDG} by a kinematic fit, and the corresponding $\chi^2$ is required to be less than 20.

\subsection{ST yields}\label{styield}

The ST $D^\mp$ candidates are formed by the combinations of $K^{\pm}\pi^{\mp}\pi^{\mp}$, $K^{\pm}\pi^{\mp}\pi^{\mp}\pi^0$, $K_S^0\pi^{\mp}$, $K_S^0\pi^{\mp}\pi^0$, $K_S^0\pi^{\pm}\pi^{\mp}\pi^{\mp}$ and $K^{\pm}K^{\mp}\pi^{\mp}$. Two variables are used to identify ST $D$ mesons: the energy difference $\Delta E$ and the beam-energy constrained mass $M_{\rm BC}$, which are defined as

\begin{align}
& \Delta E \equiv E_D - E_{\rm beam}, \\
& M_{\rm BC} \equiv \sqrt{E_{\rm beam}^2 - |\vec{\mkern1mu p}_D|^2}.
\end{align}

\noindent Here, $\vec{\mkern1mu p}_D$ and $E_D$ are the reconstructed momentum and energy of the $D$ candidate in the $e^+e^-$ c.m.\ system, and $E_{\rm beam}$ is the beam energy. Signal events are expected to peak around zero in the $\Delta E$ distribution and around the nominal $D$ mass in the $M_{\rm BC}$ distribution. In the case of multiple candidates in one event, the one with the smallest $|\Delta E|$ is chosen. Tag mode-dependent $\Delta E$ requirements as used in Ref.~\cite{klenu} are imposed on the accepted ST candidate events, as summarized in Table~\ref{ST}.

\begin{table}[hbtp]
  \centering
  \footnotesize
\caption{$\Delta E$ requirements and ST yields in data ($N_{\rm ST}$), where the uncertainties are statistical only.}\label{ST}
  \begin{spacing}{1.29}
 \begin{tabular}{L{0.30\linewidth}C{0.23\linewidth}C{0.205\linewidth}C{0.205\linewidth}}
 \hline\hline
 ST mode & $\Delta E$ (GeV) & $N_{\rm ST}$ ($D^+$)  & $N_{\rm ST}$ ($D^-$) \\
 \hline
$D^{\pm} \rightarrow K^{\mp}\pi^{\pm}\pi^{\pm}$               & $(-0.030, 0.030)$  &   412416    $\pm$  687   &   414140    $\pm$  690   \\
$D^{\pm} \rightarrow K^{\mp}\pi^{\pm}\pi^{\pm}\pi^0$          & $(-0.052, 0.039)$  &   114910    $\pm$  474   &   118246    $\pm$  479   \\
$D^{\pm} \rightarrow K_S^0\pi^{\pm}$                          & $(-0.032, 0.032)$  &   48220     $\pm$  229   &   47938     $\pm$  229   \\
$D^{\pm} \rightarrow K_S^0\pi^{\pm}\pi^0$                     & $(-0.057, 0.040)$  &   98907     $\pm$  385   &   99169     $\pm$  384   \\
$D^{\pm} \rightarrow K_S^0\pi^{\mp}\pi^{\pm}\pi^{\pm}$        & $(-0.034, 0.034)$  &   57386     $\pm$  307   &   57090     $\pm$  305   \\
$D^{\pm} \rightarrow K^{\mp}K^{\pm}\pi^{\pm}$                 & $(-0.030, 0.030)$  &   35706     $\pm$  253   &   35377     $\pm$  253   \\
\hline
\hline
 \end{tabular}
 \end{spacing}
\end{table}

\begin{figure*}[hbtp]
  \centering
  \includegraphics[width=0.3\linewidth]{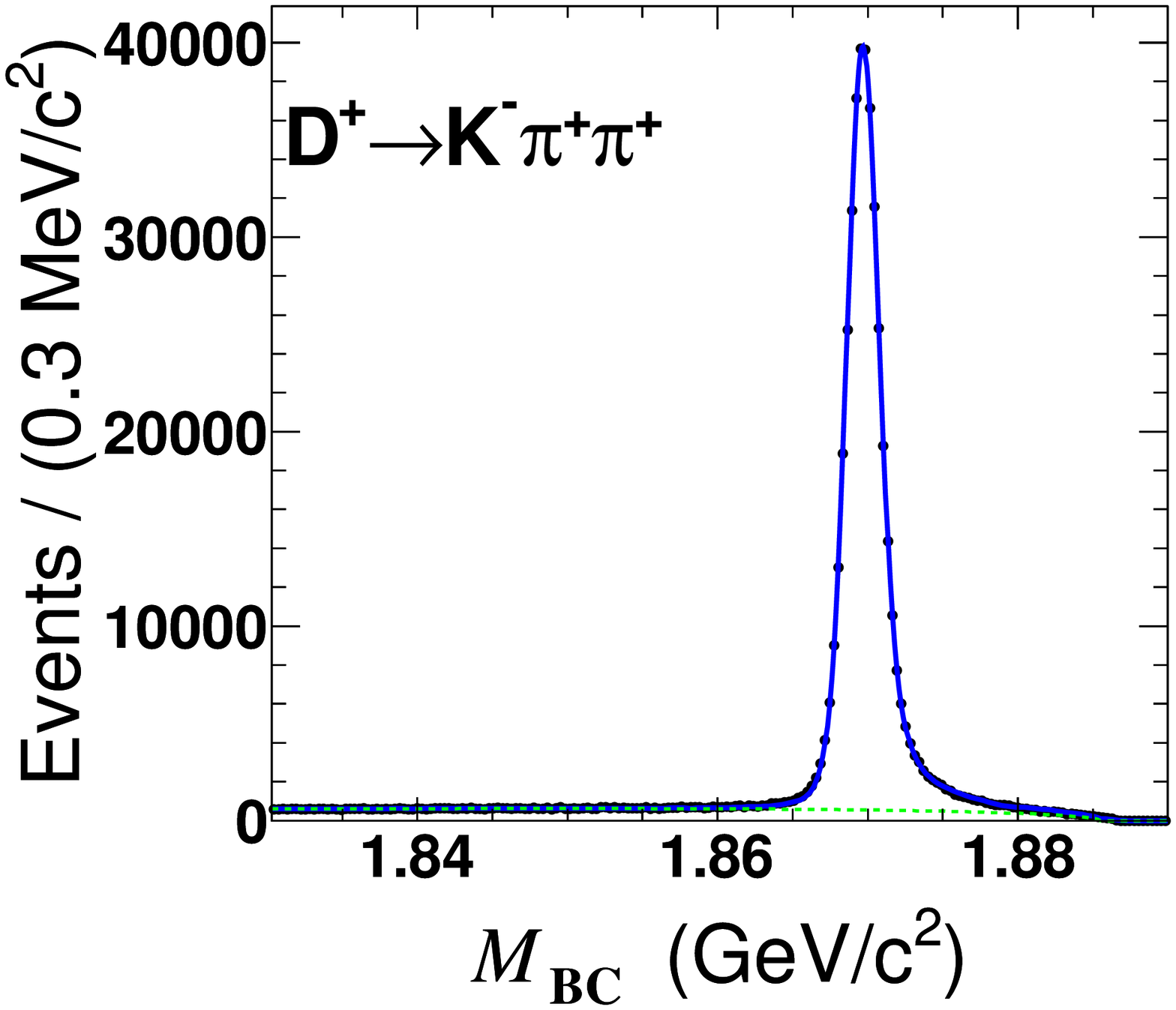}
  \includegraphics[width=0.3\linewidth]{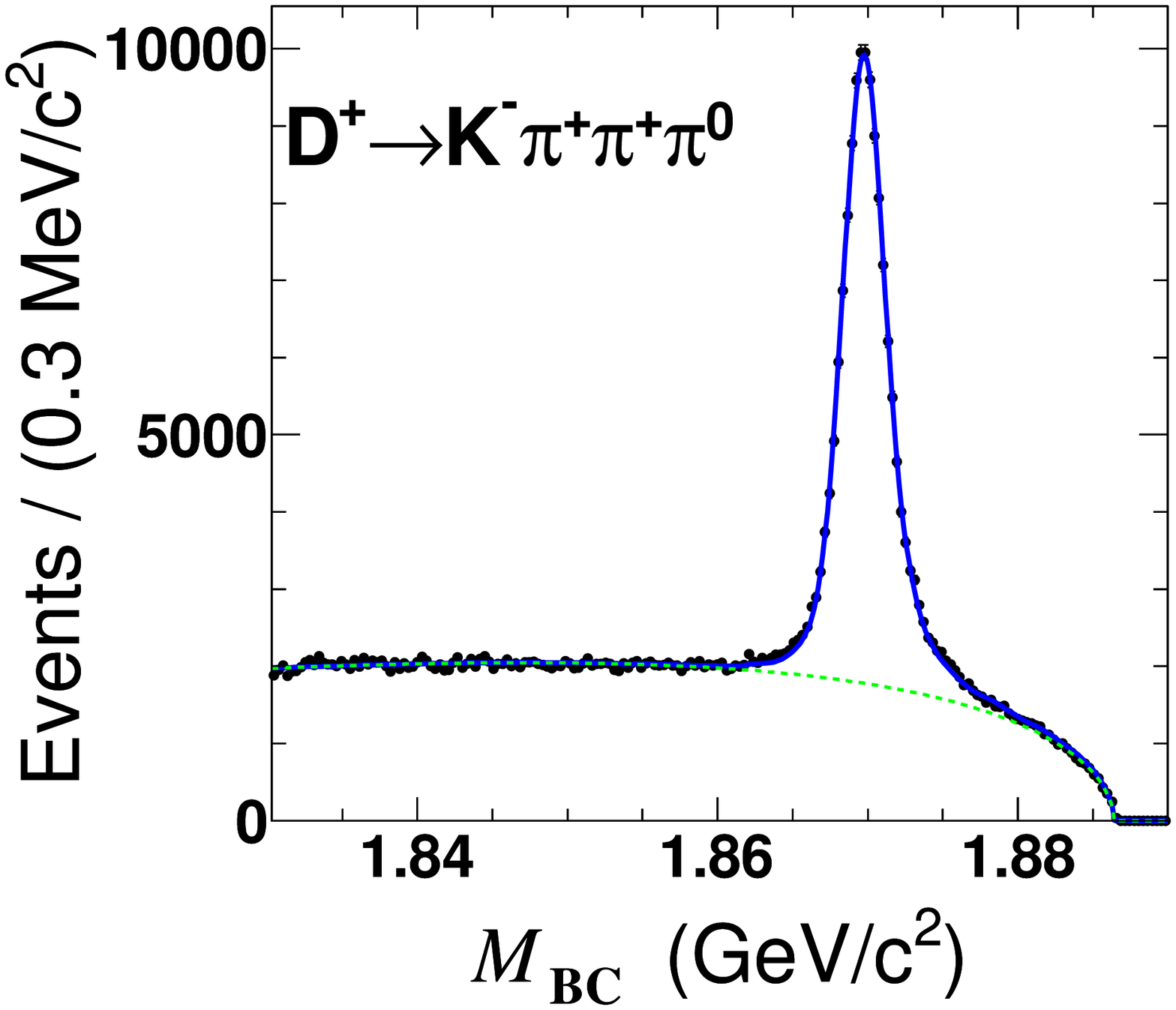}
  \includegraphics[width=0.3\linewidth]{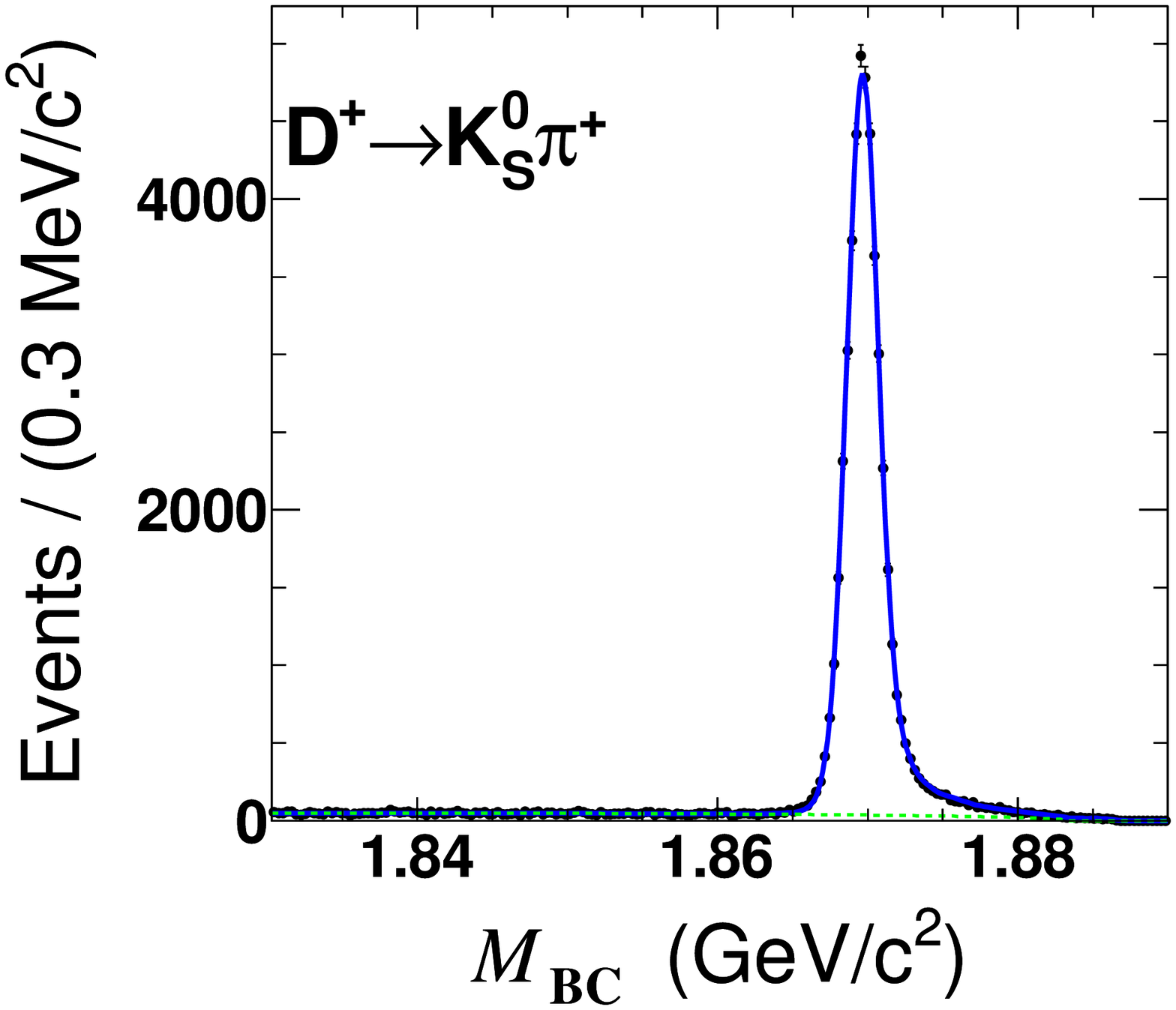}
  \includegraphics[width=0.3\linewidth]{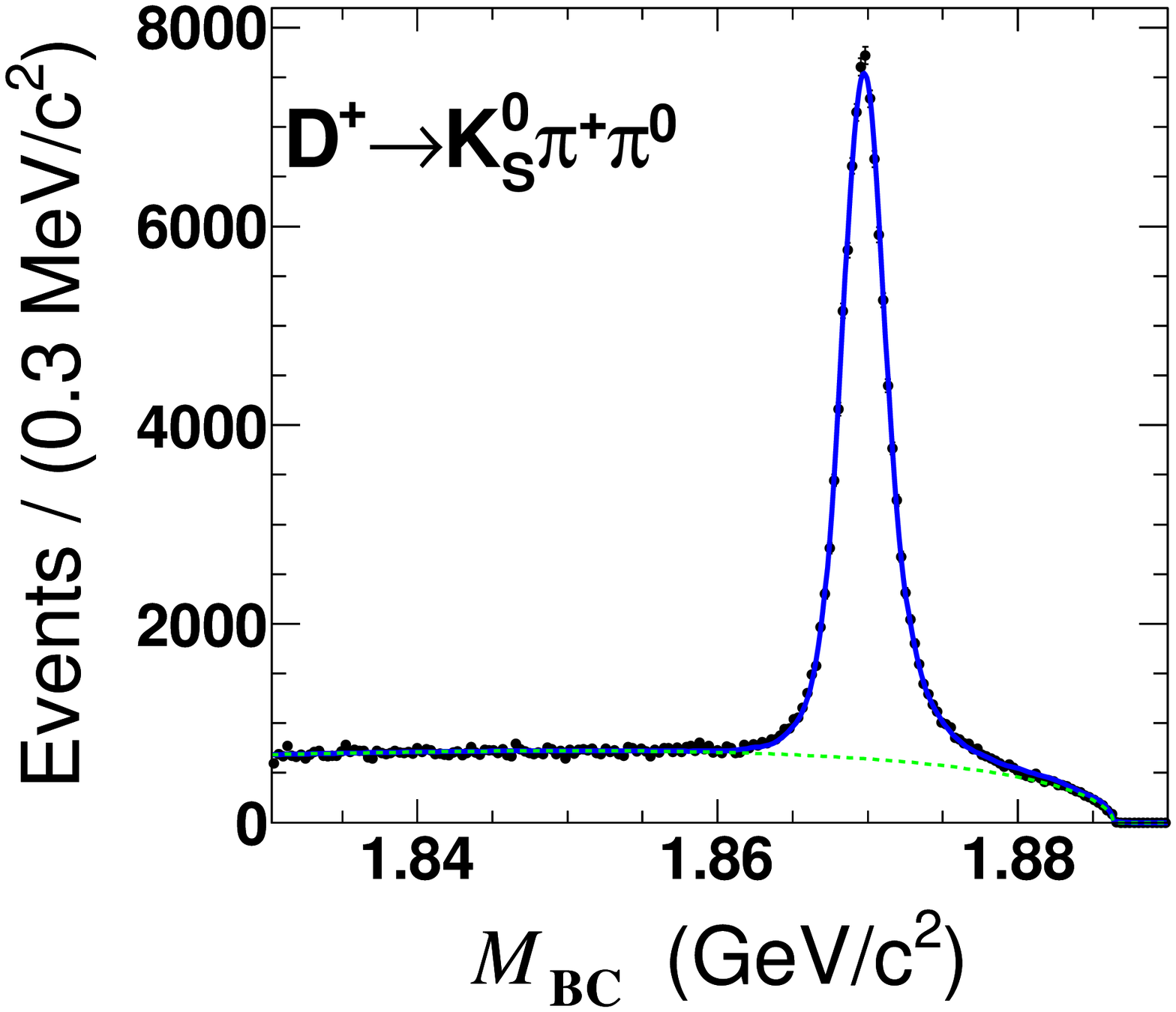}
  \includegraphics[width=0.3\linewidth]{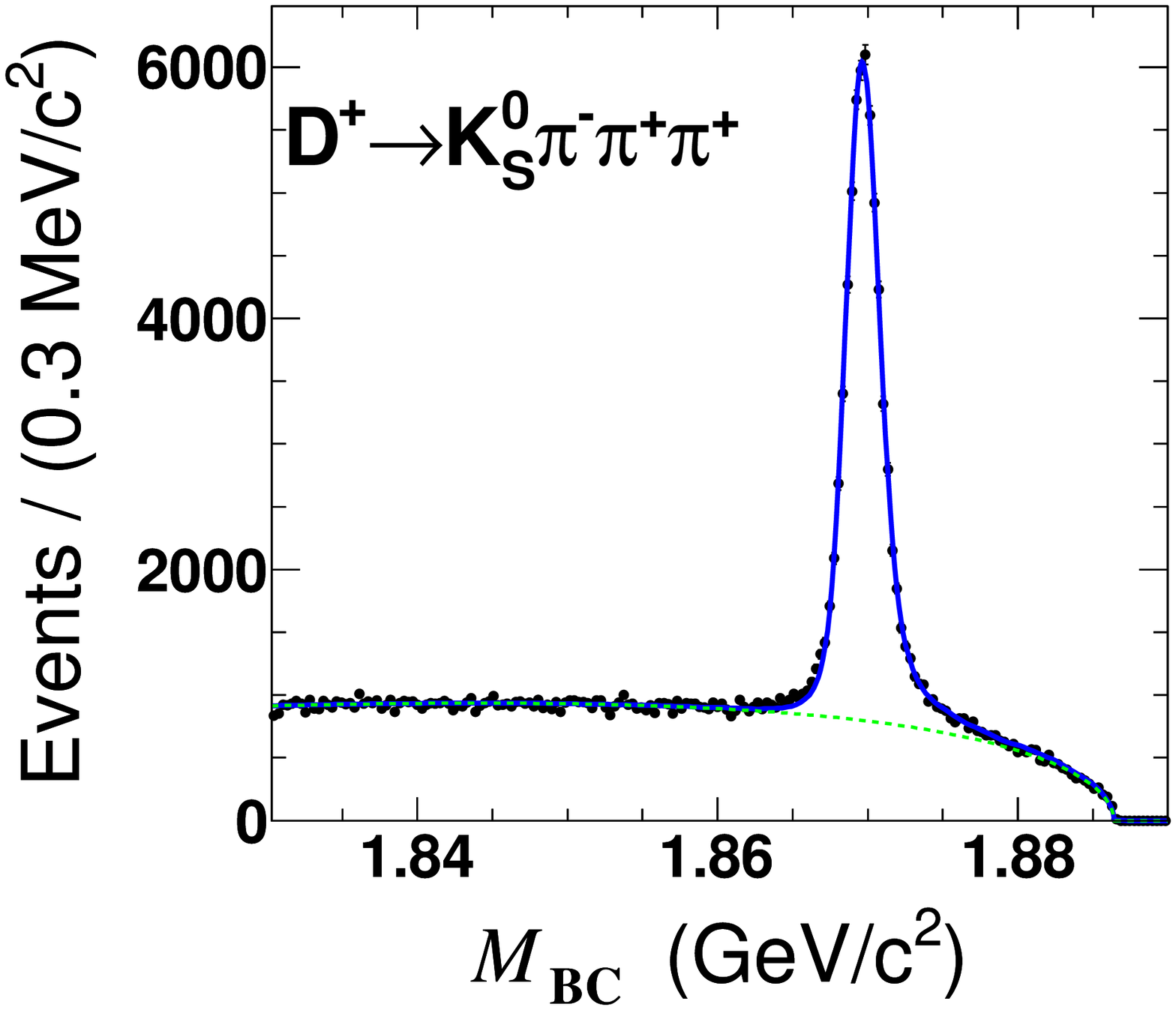}
  \includegraphics[width=0.3\linewidth]{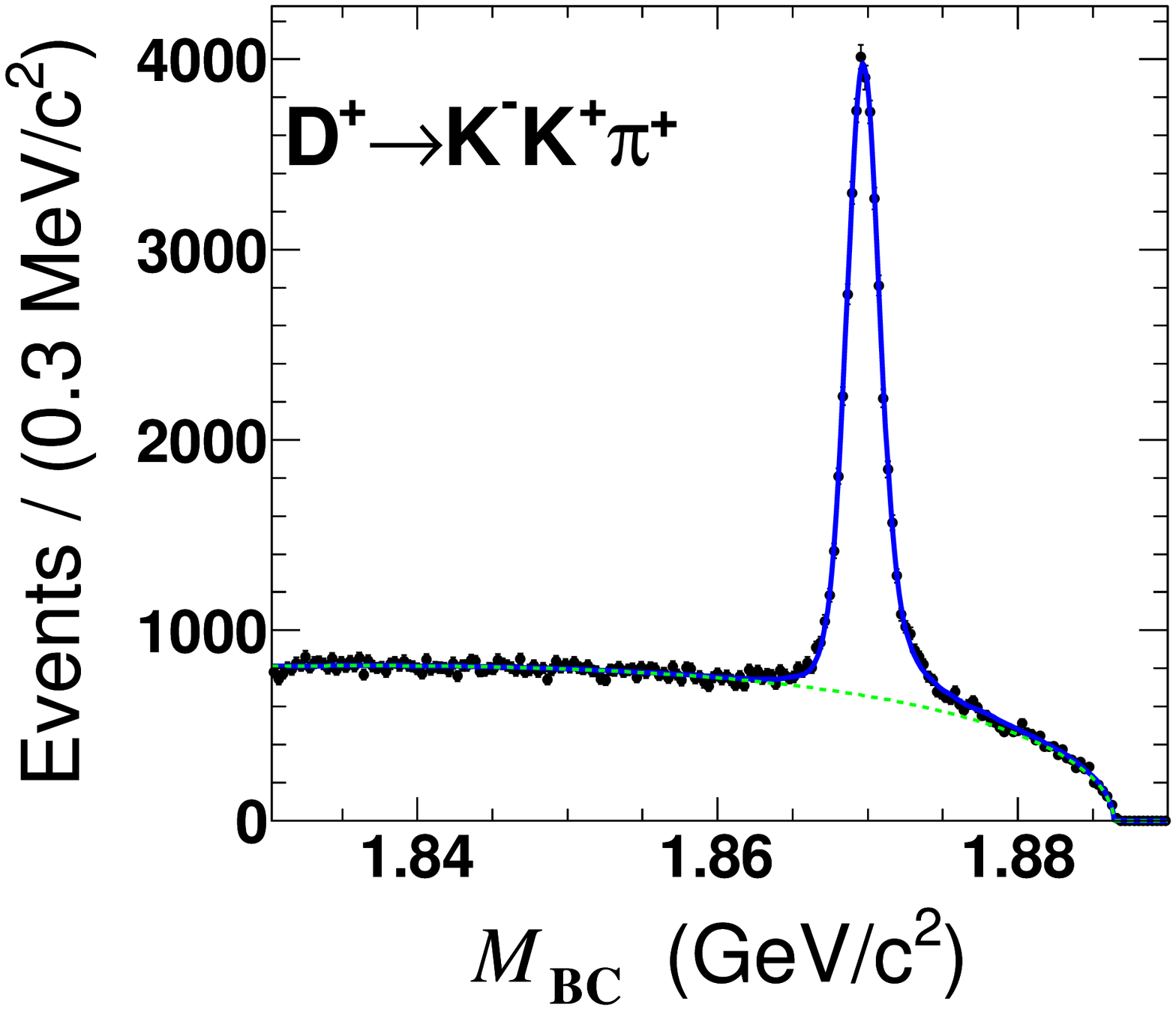}
\caption{{\color{black}{(Color online)}} Fits to the $M_{\rm BC}$ distributions of ST $D^+$ candidate events. The points with error bars are data, the green dashed curves show the fitted backgrounds, and the blue solid curves show the total fit curve.}
\label{tagDpSTmbc}
\end{figure*}

\begin{figure*}[hbtp]
  \centering
  \includegraphics[width=0.3\linewidth]{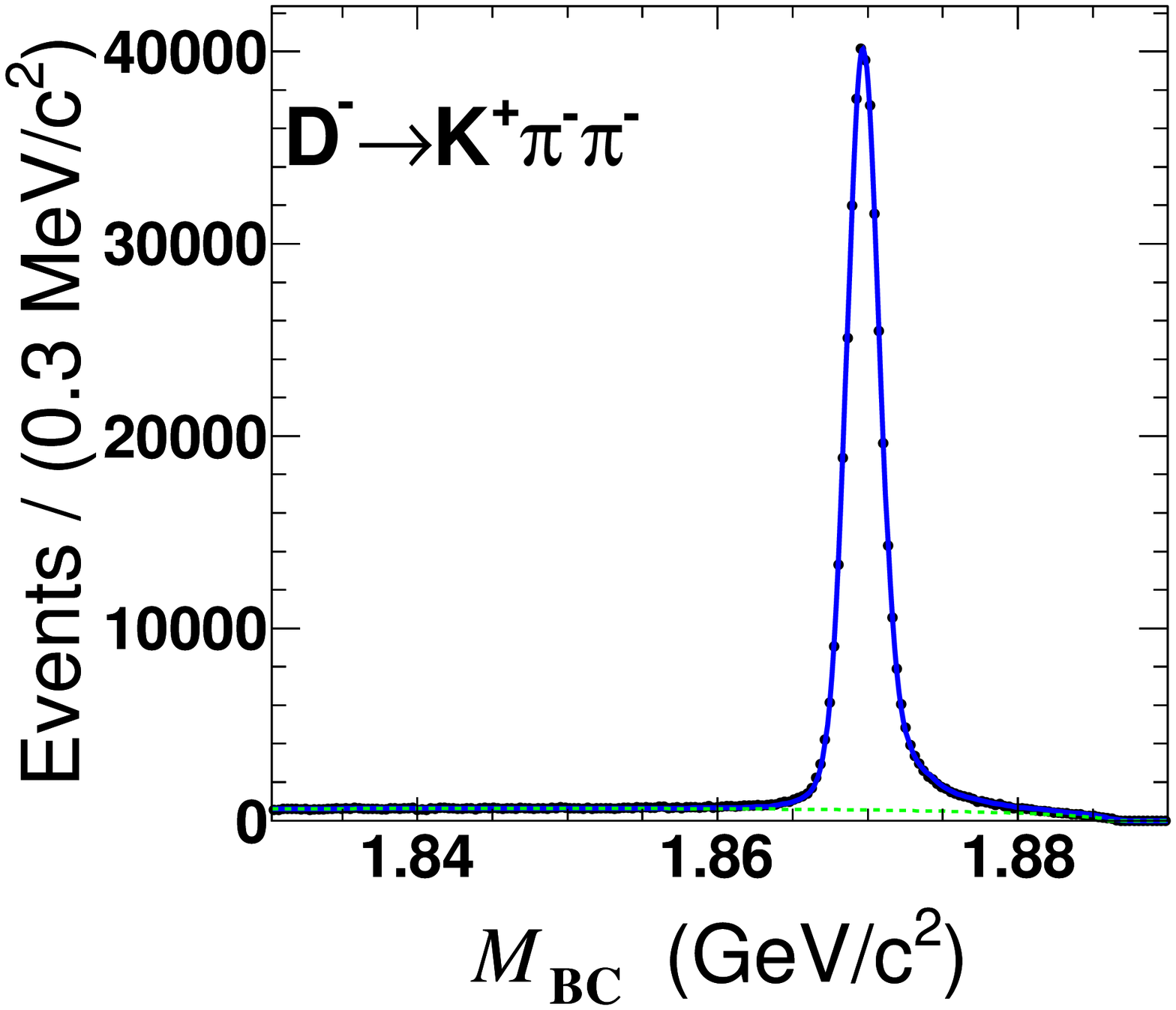}
  \includegraphics[width=0.3\linewidth]{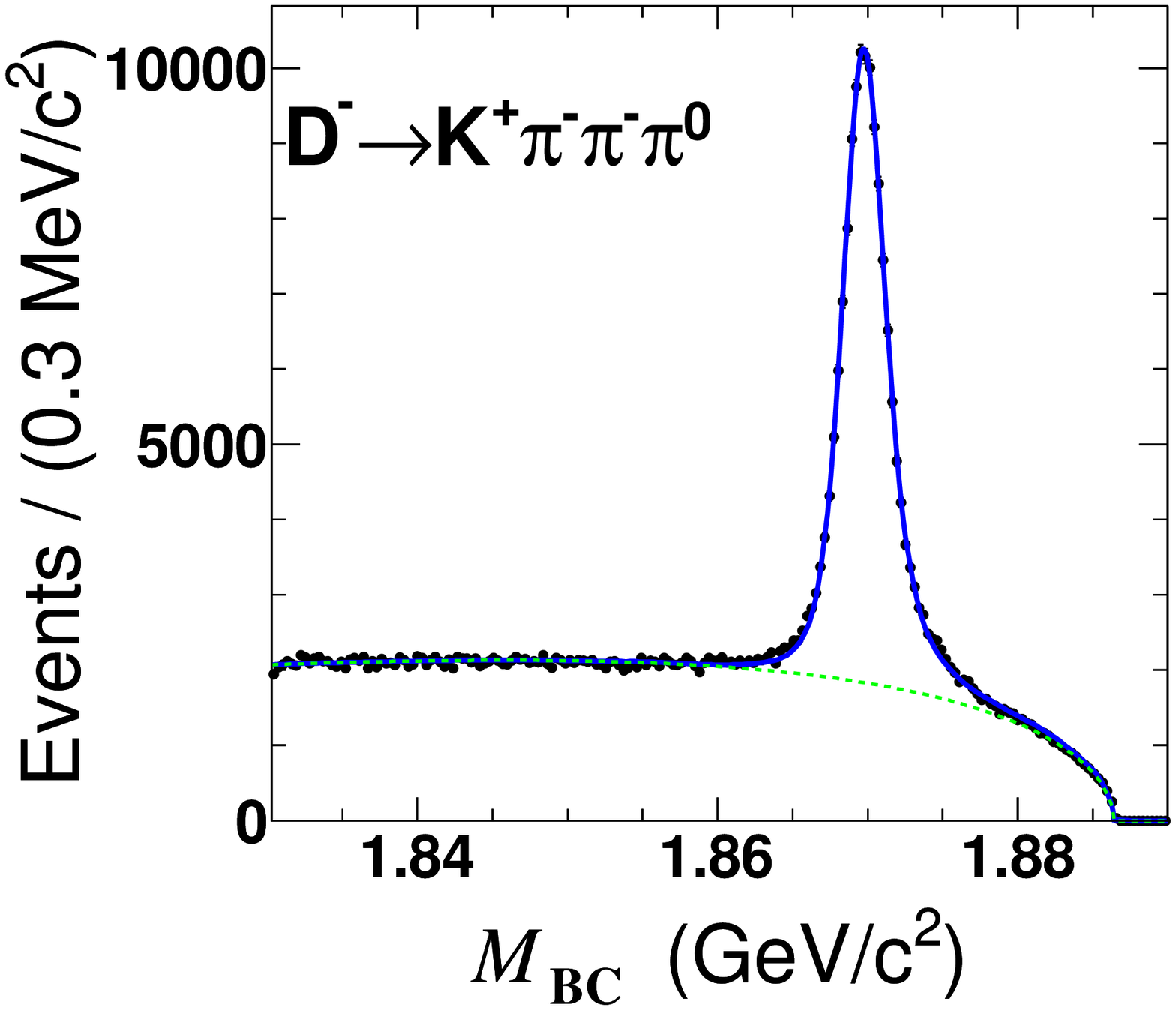}
  \includegraphics[width=0.3\linewidth]{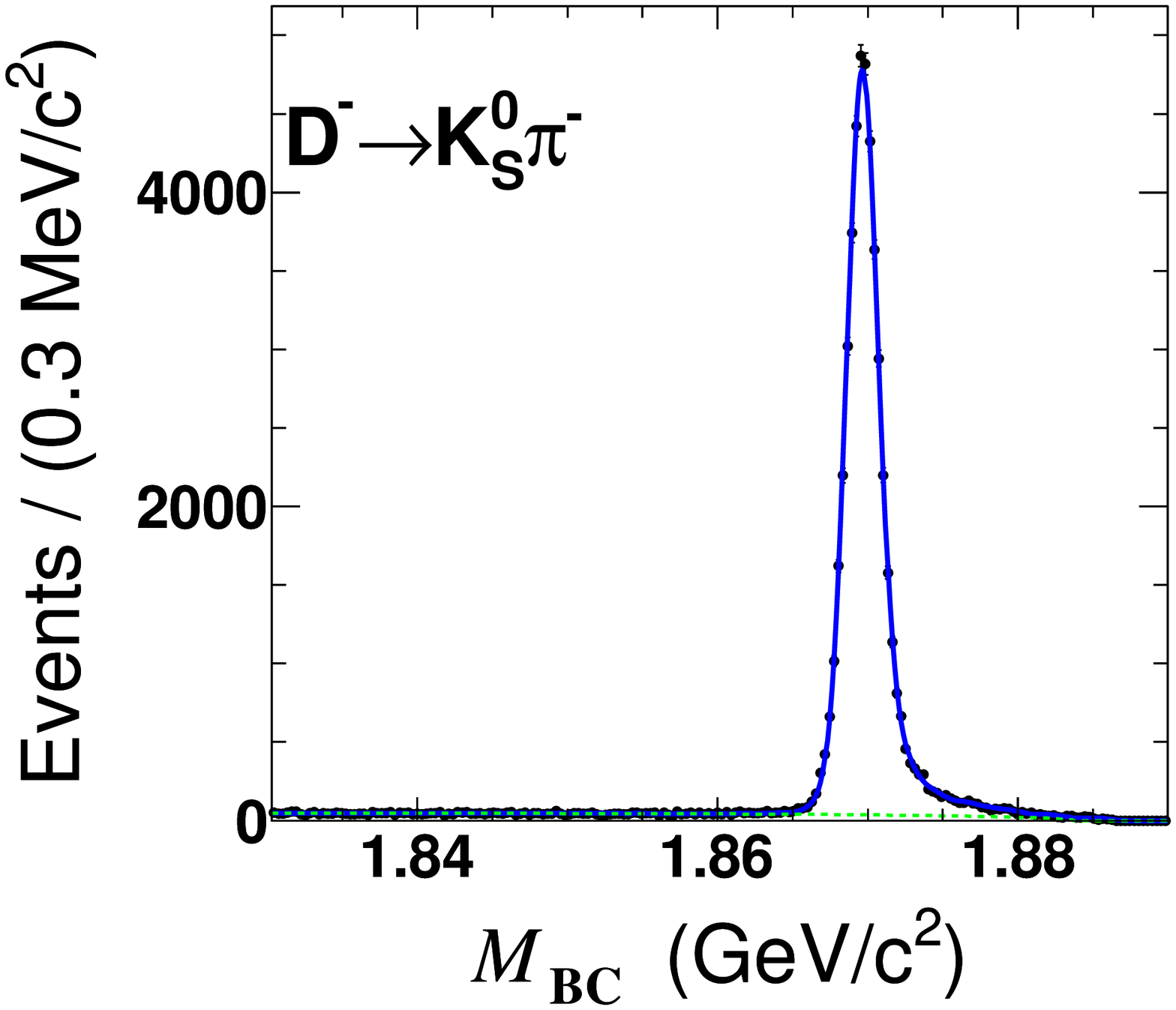}
  \includegraphics[width=0.3\linewidth]{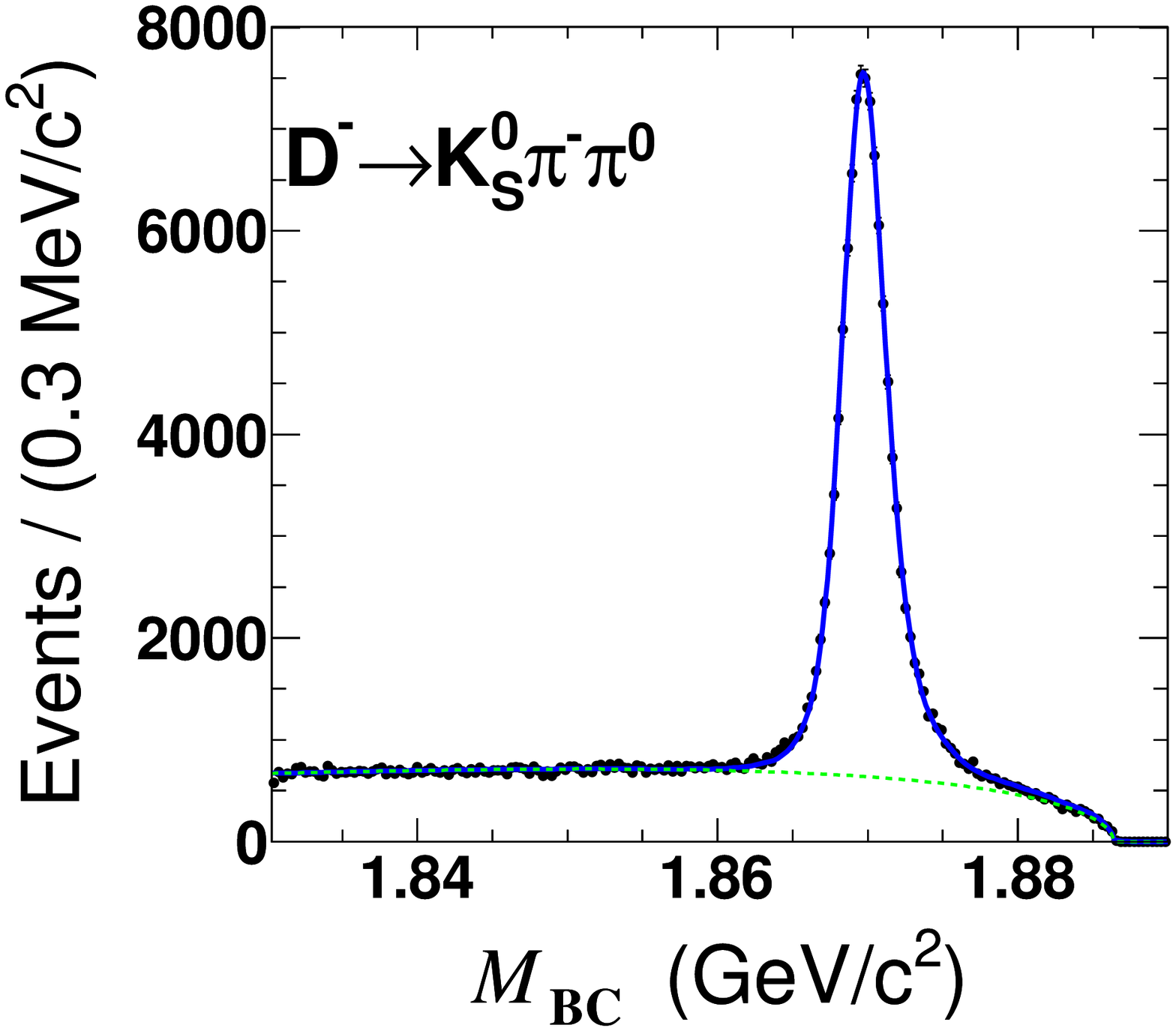}
  \includegraphics[width=0.3\linewidth]{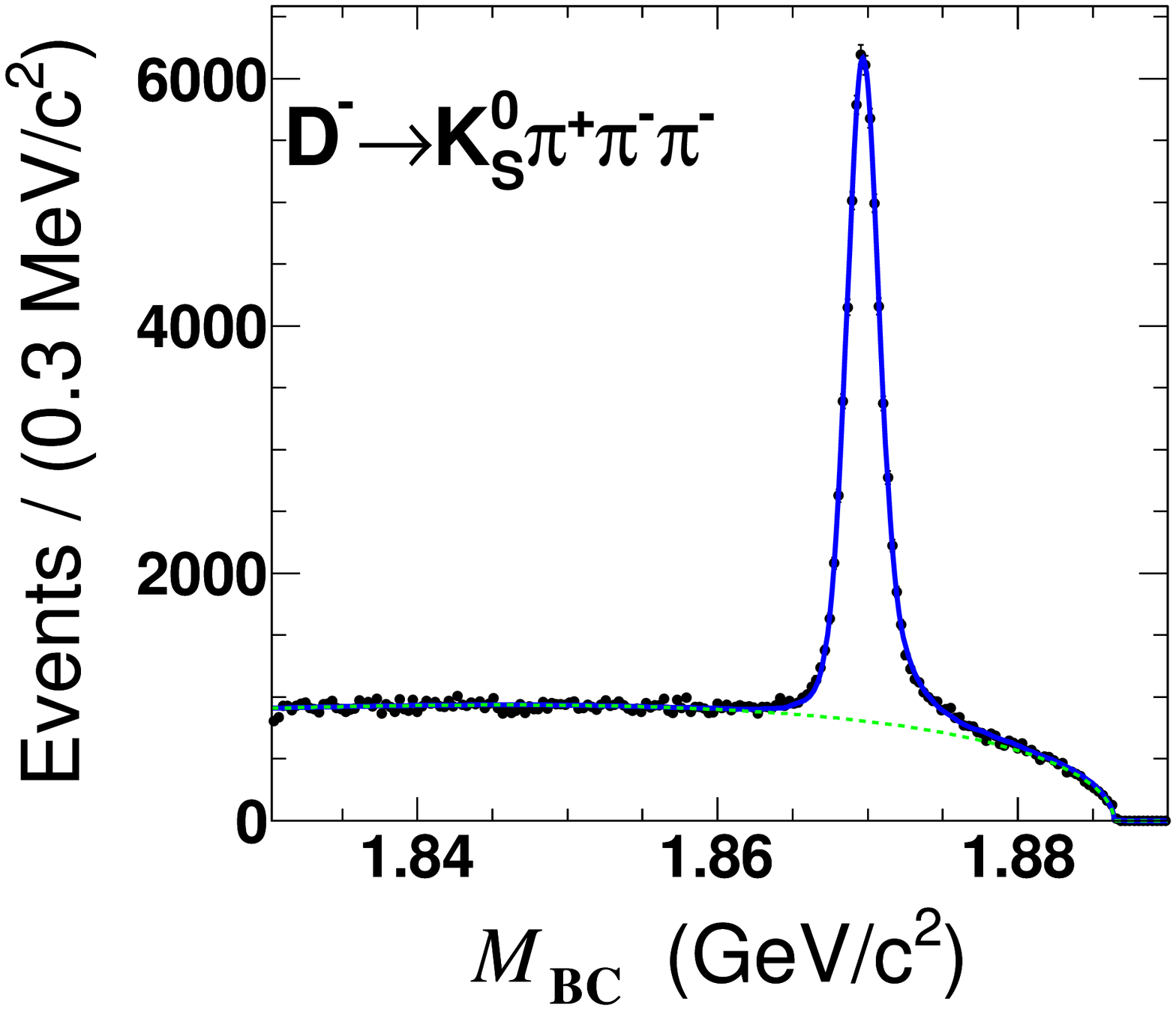}
  \includegraphics[width=0.3\linewidth]{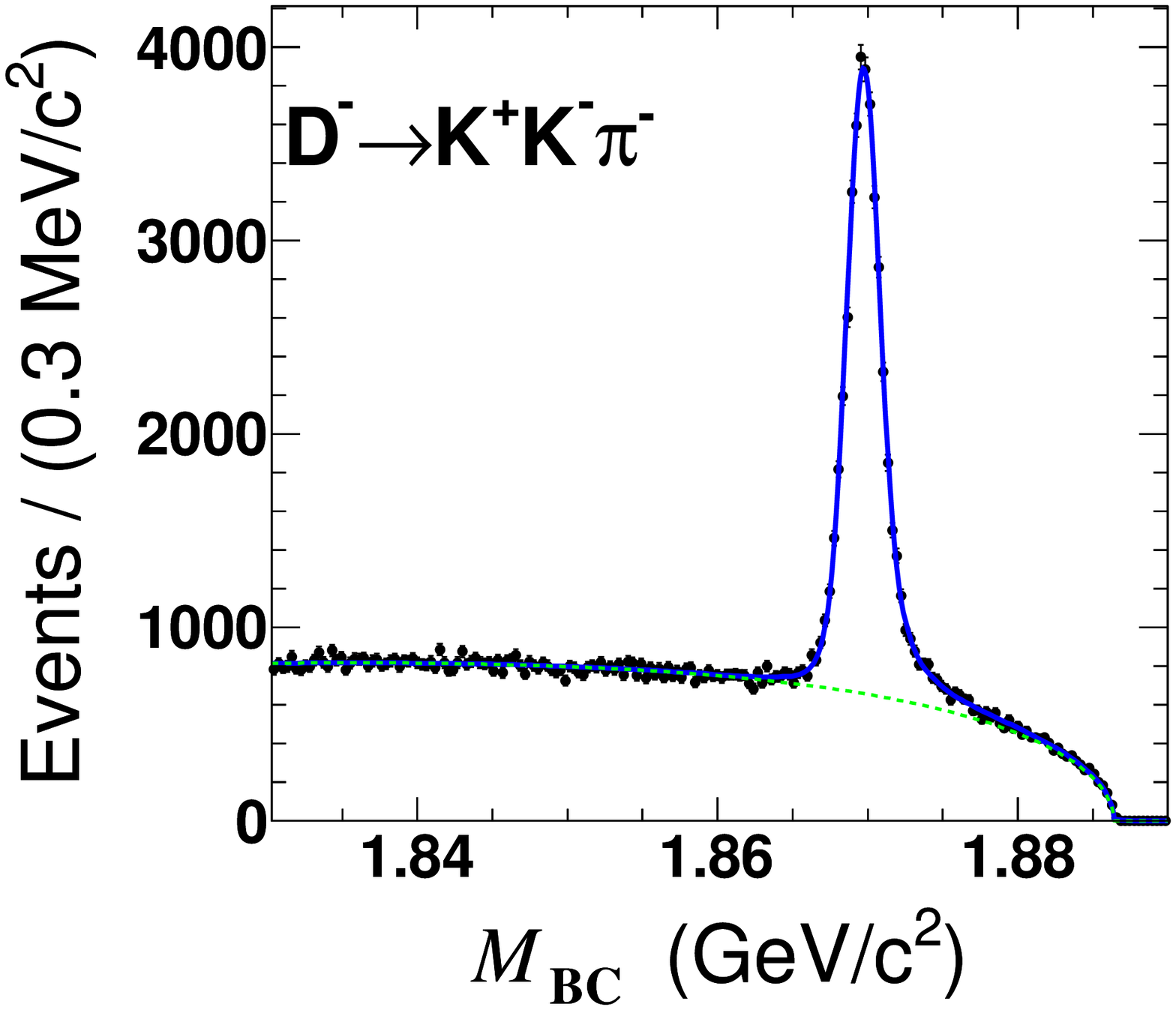}
\caption{{\color{black}{(Color online)}} Fits to the $M_{\rm BC}$ distributions of ST $D^-$ candidate events. The points with error bars are data, the green dashed curves show {\color{black}{the}} fitted backgrounds, and the blue solid curves show the total fit curve.}
\label{tagDmSTmbc}
\end{figure*}

To obtain the ST yield for each tag mode in data, a binned maximum likelihood fit is performed on the $M_{\rm BC}$ distribution, where the signal of $D$ meson is described by a MC-simulated shape and the background is modeled by an ARGUS function \cite{ARGUS}. The MC-simulated shape is convolved with a Gaussian function with free parameters to take into account the resolution difference between data and MC simulation. Figures \ref{tagDpSTmbc} and \ref{tagDmSTmbc} illustrate the resulting fits to the $M_{\rm BC}$ distributions for ST $D^+$ and $D^-$ candidate events in data, respectively. The fitted ST yields of data are presented in Table~\ref{ST}, too.

\subsection{DT yields}\label{klrec}
On the recoiling side against the ST $D^\mp$ mesons, the hadronic decays of $D^{\pm}\rightarrow K_{S,L}^0 K^{\pm}(\pi^0)$ are selected using the remaining tracks and neutral clusters. The charged kaon is required to have the same charge as the signal $D$ meson candidate. To suppress backgrounds, no extra good charged track is allowed in the DT candidate events. The signal $D^\pm$ candidates are also identified with the energy difference and the beam energy constrained mass.
In the following, the energy difference and the beam-energy constrained mass of the particle combination for the ST/signal side are denoted as $\Delta E^{\rm tag/sig}$ and $M_{\rm BC}^{\rm tag/sig}$, respectively. In each event, if there are multiple signal candidates for $D^{\pm}\rightarrow K_S^0 K^{\pm}(\pi^0)$, the one with the smallest $|\Delta E^{\rm sig}|$ is selected. The $\Delta E^{\rm sig}$ is required to be within $(-0.031,  0.031)$~GeV and $(-0.057,  0.040)$~GeV for $D^\pm\to K_S^0 K^{\pm}$ and $D^\pm\to K_S^0 K^{\pm}\pi^0$, respectively.

Due to its long lifetime, very few $K_L^0$ decay in the MDC. However, most $K_L^0$ will interact in the material of the EMC, which gives their position but no reliable measurement of their energy. Thus, to select the candidates of $D^{\pm}\rightarrow K_L^0 K^{\pm}(\pi^0)$, the momentum direction of the $K_L^0$ particle is inferred by the position of a shower in the EMC, and a kinematic fit imposing momentum and energy conversation for the observed particles and a missing $K_L^0$ particle is performed to select the signal, where the $K_L^0$ particle is of known mass and momentum direction, but of unknown momentum magnitude. We perform the kinematic fit individually for all shower candidates in the EMC that are not used in the ST side and do not form a $\pi^0$ candidate with any other shower candidate with invariant mass within $(0.110, 0.155)$~GeV/$c^2$~\cite{klenu}. The candidate with the minimal chi-square of the kinematic fit ($\chi^2_{K^0_L}$) is selected. To minimize the correlation between $M_{\rm BC}^{\rm tag}$ and $M_{\rm BC}^{\rm sig}$, the momentum of the $K_L^0$ candidate is not taken from the kinematic fit, but inferred by constraining $\Delta E^{\rm sig}$ to be zero. In order to suppress backgrounds due to cluster candidates produced mainly from electronics noise, the energy of the $K_L^0$ shower in the EMC is required to be greater than 0.1~GeV. Finally, DT candidate events are imposed with the optimized, and ST and signal mode dependent $\chi_{K_L^0}^2$ requirements, as summarized in Table~\ref{Tchisq}.

\begin{table}[hbtp]
  \centering
  \footnotesize
  \caption{Requirements on $\chi_{K_L^0}^2$ for DT signal  events.}\label{Tchisq}
  \begin{spacing}{1.29}
   \begin{tabular}{L{0.3\linewidth}C{0.32\linewidth}C{0.32\linewidth}}
 \hline\hline
  ST mode    &  $D^{\pm} \rightarrow K_L^0 K^{\pm}$   &  $D^{\pm} \rightarrow K_L^0 K^{\pm}\pi^0$ \\
\hline
$D^{\mp} \rightarrow K^{\pm}\pi^{\mp}\pi^{\mp}$               &  80   &   80 \\
$D^{\mp} \rightarrow K^{\pm}\pi^{\mp}\pi^{\mp}\pi^0$          &  50   &   40 \\
$D^{\mp} \rightarrow K_S^0\pi^{\mp}$                          &  80   &   50 \\
$D^{\mp} \rightarrow K_S^0\pi^{\mp}\pi^0$                     &  40   &   25 \\
$D^{\mp} \rightarrow K_S^0\pi^{\mp}\pi^{\mp}\pi^{\pm}$        &  40   &   30 \\
$D^{\mp} \rightarrow K^{\pm}K^{\mp}\pi^{\mp}$                 &  40   &   40 \\
\hline
\hline
 \end{tabular}
 \end{spacing}
\end{table}

\begin{figure}[hbtp]
\centering
\normalsize
  \begin{overpic}[width=0.95\linewidth]{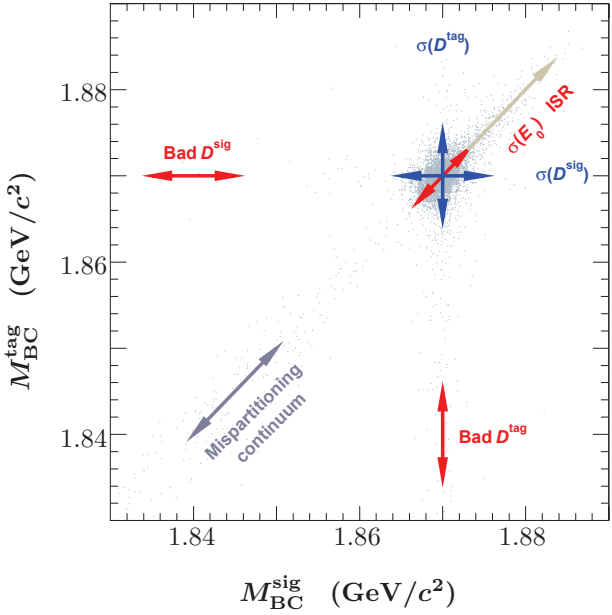}
  \put(3,37){\begin{rotate}{90}$\bm{M_{\rm BC}^{\rm tag}}$ \; $\bm{({\rm GeV/}c^2)}$ \end{rotate}}
  \put(37,3){$\bm{M_{\rm BC}^{\rm sig}}$ \; $\bm{({\rm GeV/}c^2)}$}
  \put(8,28){1.84} \put(8,56){1.86} \put(8,84){1.88}
  \put(25.5,12){1.84} \put(52.25,12){1.86} \put(79,12){1.88}
  \end{overpic}
  \caption{Illustration of the scatter plot of $M_{\rm BC}^{\rm tag}$ versus $M_{\rm BC}^{\rm sig}$ from the DT candidate evens of $D^+ \to K^0_S\pi^+$, summed over six ST modes.}
\label{twoDfit}
\end{figure}

Figure \ref{twoDfit} illustrates the distribution of $M_{\rm BC}^{\rm tag}$ versus $M_{\rm BC}^{\rm sig}$ for the DT candidate events of $D^+ \rightarrow K_S^0 \pi^+$, summed over the six ST modes. Candidate signal events concentrate around the intersection of $M_{\rm BC}^{\rm tag} = M_{\rm BC}^{\rm sig} = M_{D^+}$, where $M_{D^+}$ is the nominal $D^+$ mass \cite{PDG}. Candidate events with one correctly reconstructed and one incorrectly reconstructed  $D$ meson are spread along the vertical band with $M_{\rm BC}^{\rm sig} = M_{D^+}$ or horizontal band with $M_{\rm BC}^{\rm tag} = M_{D^+}$, respectively (named BKGI thereafter). Other candidate events, smeared along the diagonal, are mainly from the continuum process $e^+e^- \rightarrow q\bar q$ (named BKGII thereafter). To determine the DT signal yield, we perform an unbinned two-dimensional (2D) maximum likelihood fit on the distribution of $M_{\rm BC}^{\rm tag}$ versus $M_{\rm BC}^{\rm sig}$ of the selected events. In the fit, the probability density functions for the signal, BKGI and BKGII components are constructed as follows:

\begin{itemize}
\item Signal: $a(x, y)\otimes g(x; x_0, \sigma_{x_0})\otimes g(y; y_0, \sigma_{y_0})$,
\item BKGI:   $b(x, y)$,
\item BKGII:  $c((x+y)/\sqrt{2}; m_0, \xi, \rho)\cdot(G~g((x-y)/\sqrt{2}; z, \sigma_{z1}) + (1-G)~g((x-y)/\sqrt{2}; z, \sigma_{z2}))$,
\end{itemize}

\noindent where $x(y)$ denotes $M_{\rm BC}^{\rm sig(tag)}$. The signal is described with a MC-simulated shape $a(x,y)$ convolved with two independent Gaussian functions $g(x;x_0,\sigma_{x_0})$ and $g(y;y_0,\sigma_{y_0})$ representing the resolution difference between data and MC simulation in the variables $M_{\rm BC}^{\rm sig}$ and $M_{\rm BC}^{\rm tag}$, respectively. The parameters of the Gaussian functions $x_0,\sigma_{x_0}, y_0,\sigma_{y_0}$ are determined by performing 1D fits on the $M_{\rm BC}^{\rm sig}$ and $M_{\rm BC}^{\rm tag}$ distributions of data, individually.

The shape of BKGI $b(x,y)$ is determined from the generic $D \bar D$ MC sample. In particular, in the studies of $D^+\rightarrow K_L^0 K^+(\pi^0)$, irreducible and peaking backgrounds come mainly from $D^+\rightarrow K_S^0 K^+(\pi^0)$ with $K_S^0 \rightarrow \pi^0\pi^0$. Since their shape is too similar to be separated from the signal in the fit, their size and shape are fixed. To take into account possible differences between data and MC simulation, both shapes and magnitudes of the $D^+\rightarrow K_S^0 K^+(\pi^0)$ background events are re-estimated as follows. The background shapes are determined by imposing the same selection criteria as for data on the MC samples of $D^+\rightarrow K_S^0 K^+(\pi^0)$ with $K_S^0$ decaying inclusively. The background magnitudes are estimated by using the samples of $D^+\rightarrow K_S^0 K^+(\pi^0)$ with $K_S^0 \rightarrow \pi^0\pi^0$ selected from data and MC samples, from which the event yields $N_{K_S^0}^{\rm DT}$ and $N_{K_S^0}^{\rm MC}$ are determined individually. We also apply the selection criteria of $D^+\rightarrow K_L^0 K^+(\pi^0)$ on the same MC samples of $D^+\rightarrow K_S^0 K^+(\pi^0)$ with $K_S^0$ decaying inclusively, selecting $N_{K_L^0}^{\rm MC}$ events. The number of background events is then estimated by $N_{K_S^0}^{\rm DT}\cdot {N_{K_L^0}^{\rm MC}}/{N_{K_S^0}^{\rm MC}}$.

The shape of BKGII is described with an ARGUS function~\cite{ARGUS}, $c((x+y)/\sqrt{2}; m_0, \xi, \rho) = A(x+y)/\sqrt{2}(1 - \frac {(x+y)^2}{2 m_0^2})^{\rho} \cdot e^{\xi(1-\frac {(x+y)^2}{2m_0^2})}$, multiplied by a double Gaussian function. The parameters $A$ and $\xi$ of the ARGUS function are obtained by fitting the $(x+y)/\sqrt{2}$ distribution with fixed values $\rho=0.5$ and $m_0=1.8865$~GeV/$c^2$, and the parameters $z$, $\sigma_{z1}$, $\sigma_{z2}$ and $G$ of the double Gaussian function are obtained by a fit to the $(x-y)/\sqrt{2}$ distribution of data.

\begin{figure*}[hbtp]
  \centering
  \begin{overpic}[width=0.23\linewidth]{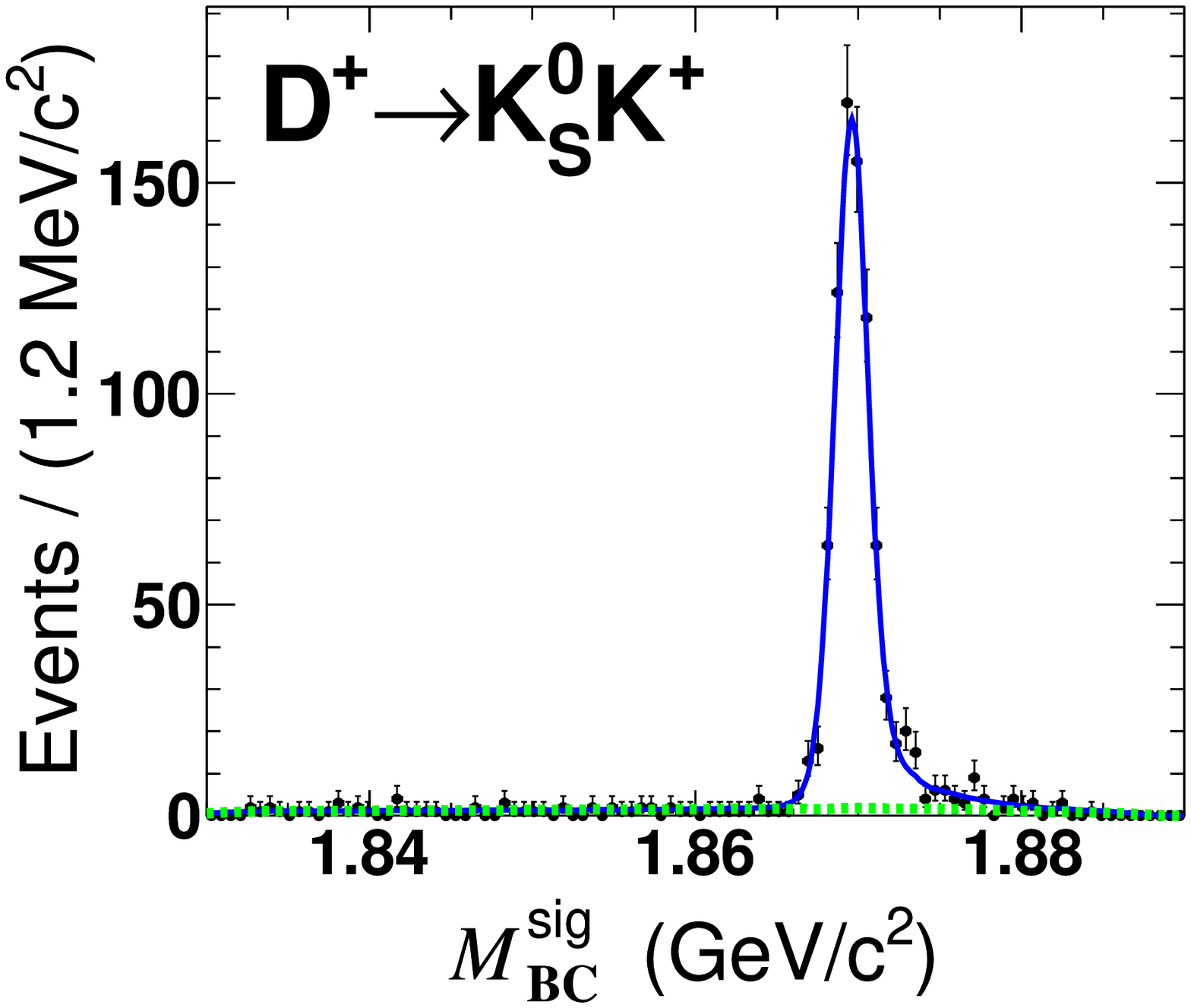}
  \end{overpic}
  \begin{overpic}[width=0.23\linewidth]{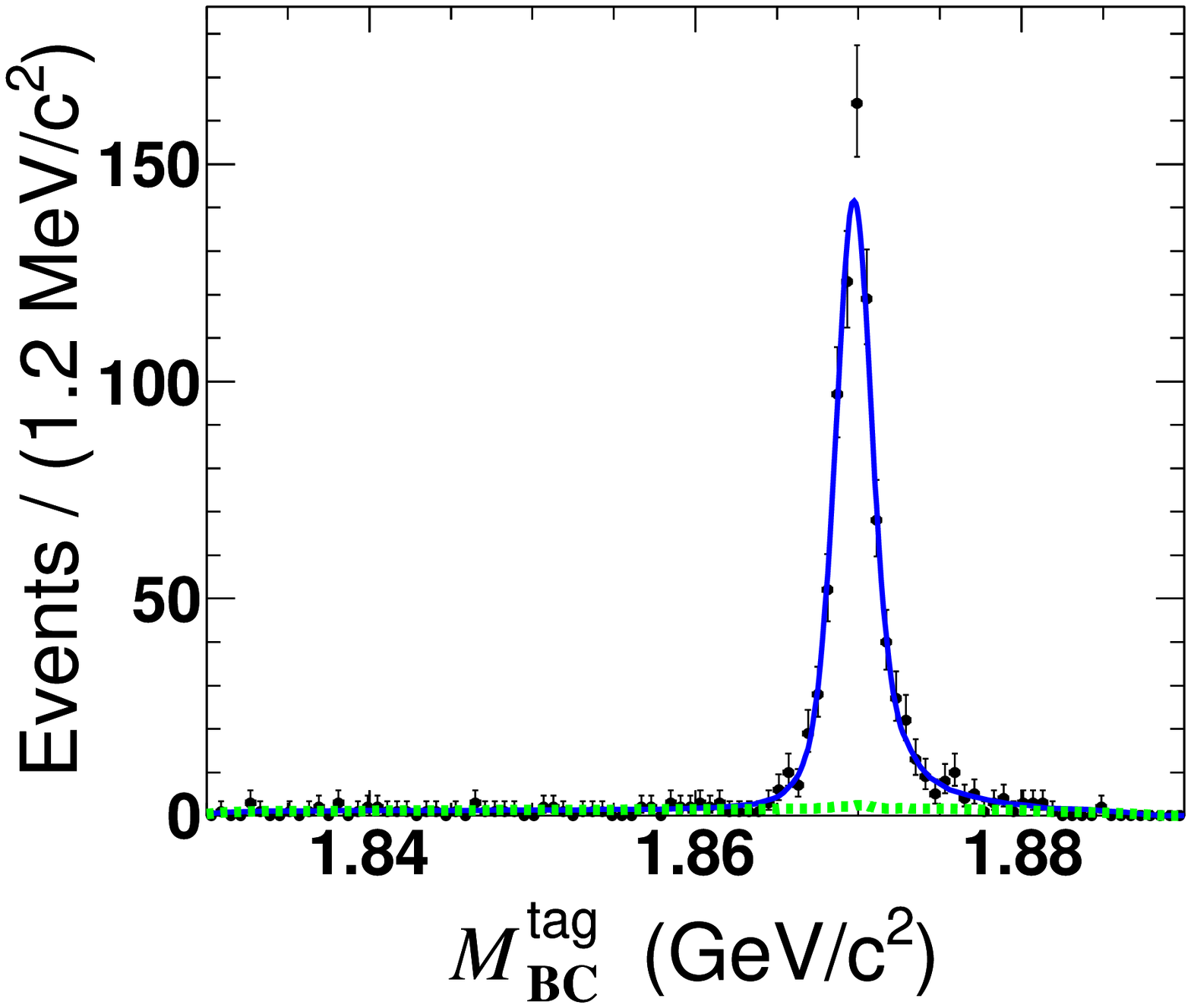}
  \end{overpic}
  \begin{overpic}[width=0.23\linewidth]{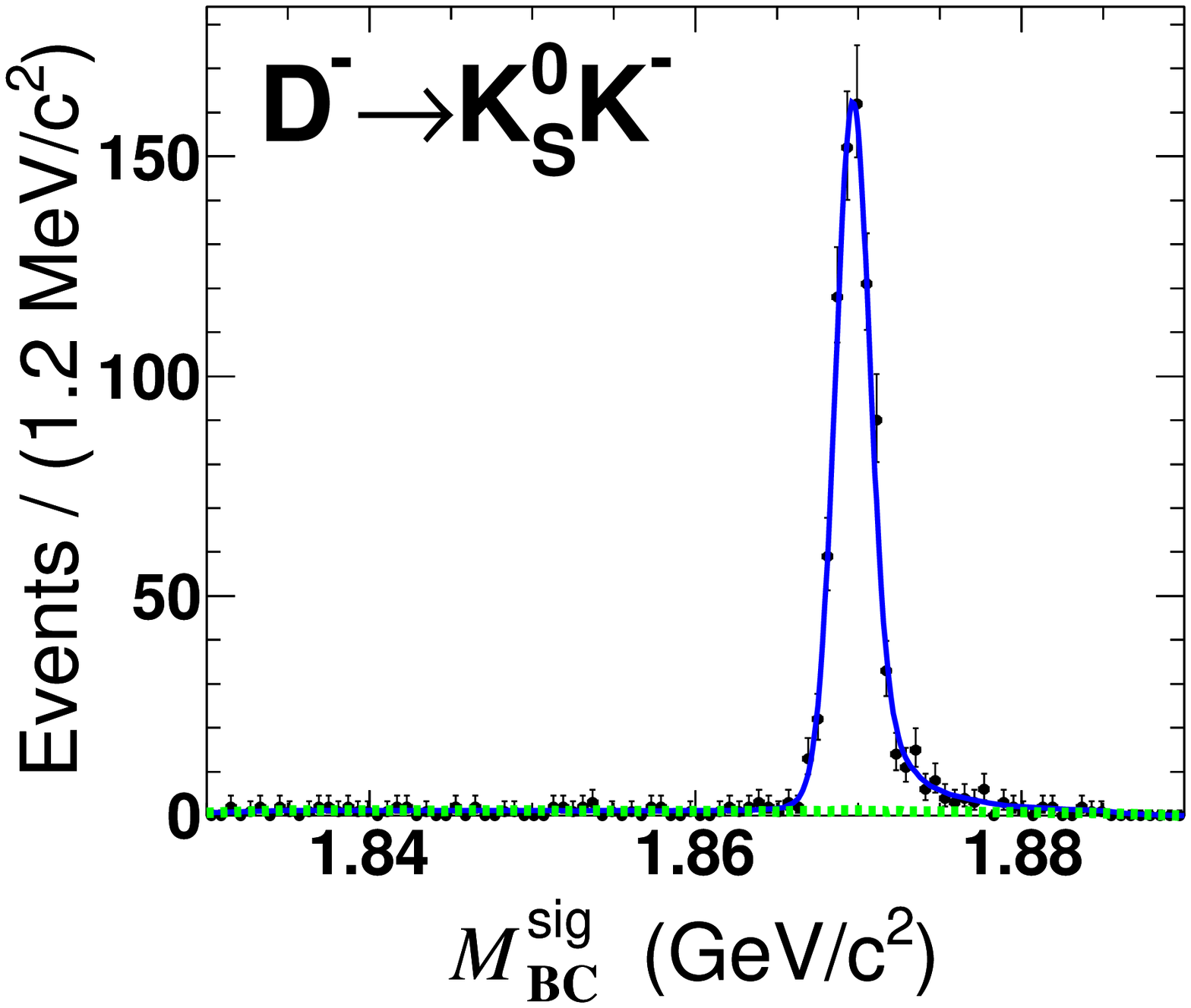}
  \end{overpic}
  \begin{overpic}[width=0.23\linewidth]{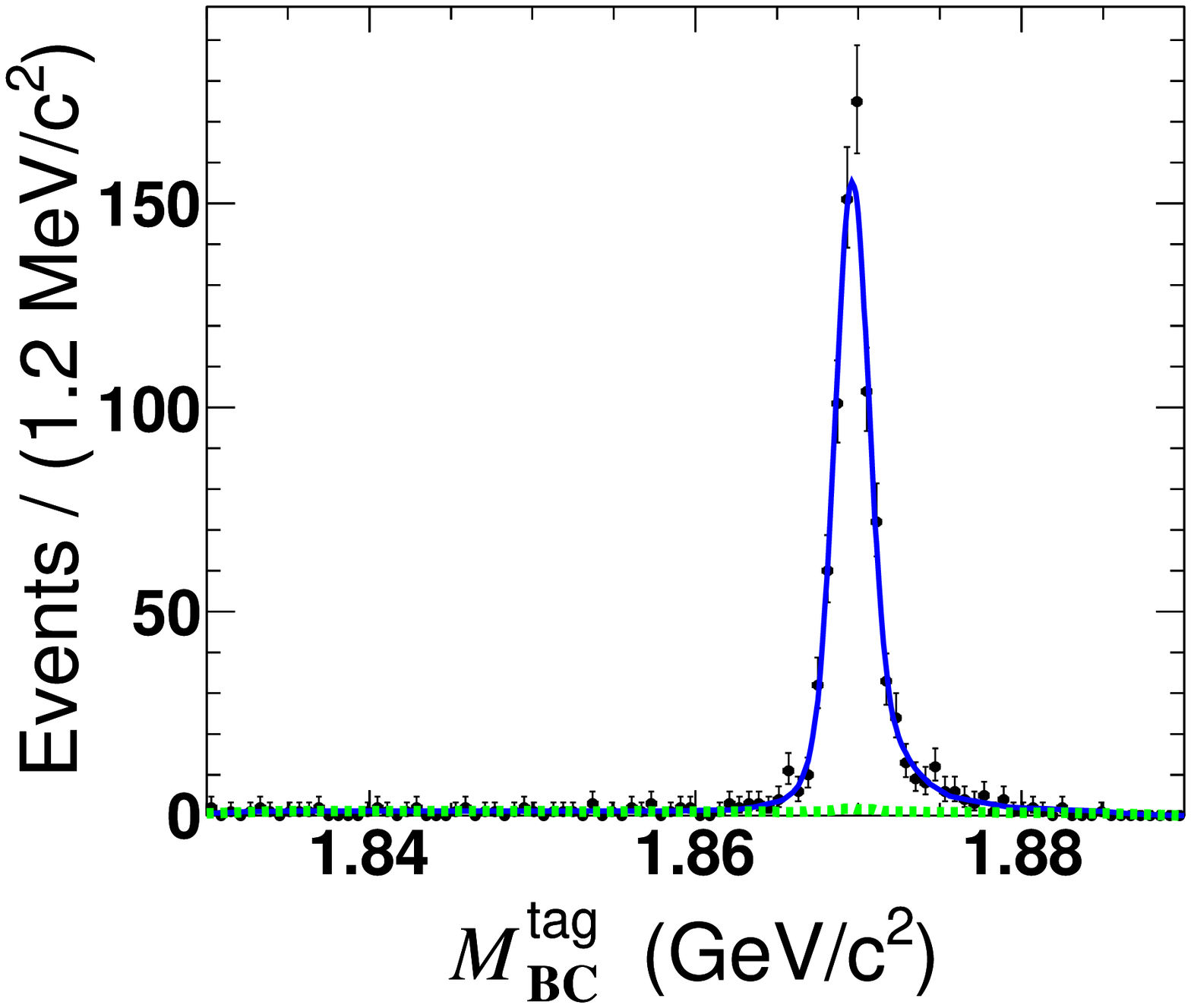}
  \end{overpic}
  \begin{overpic}[width=0.23\linewidth]{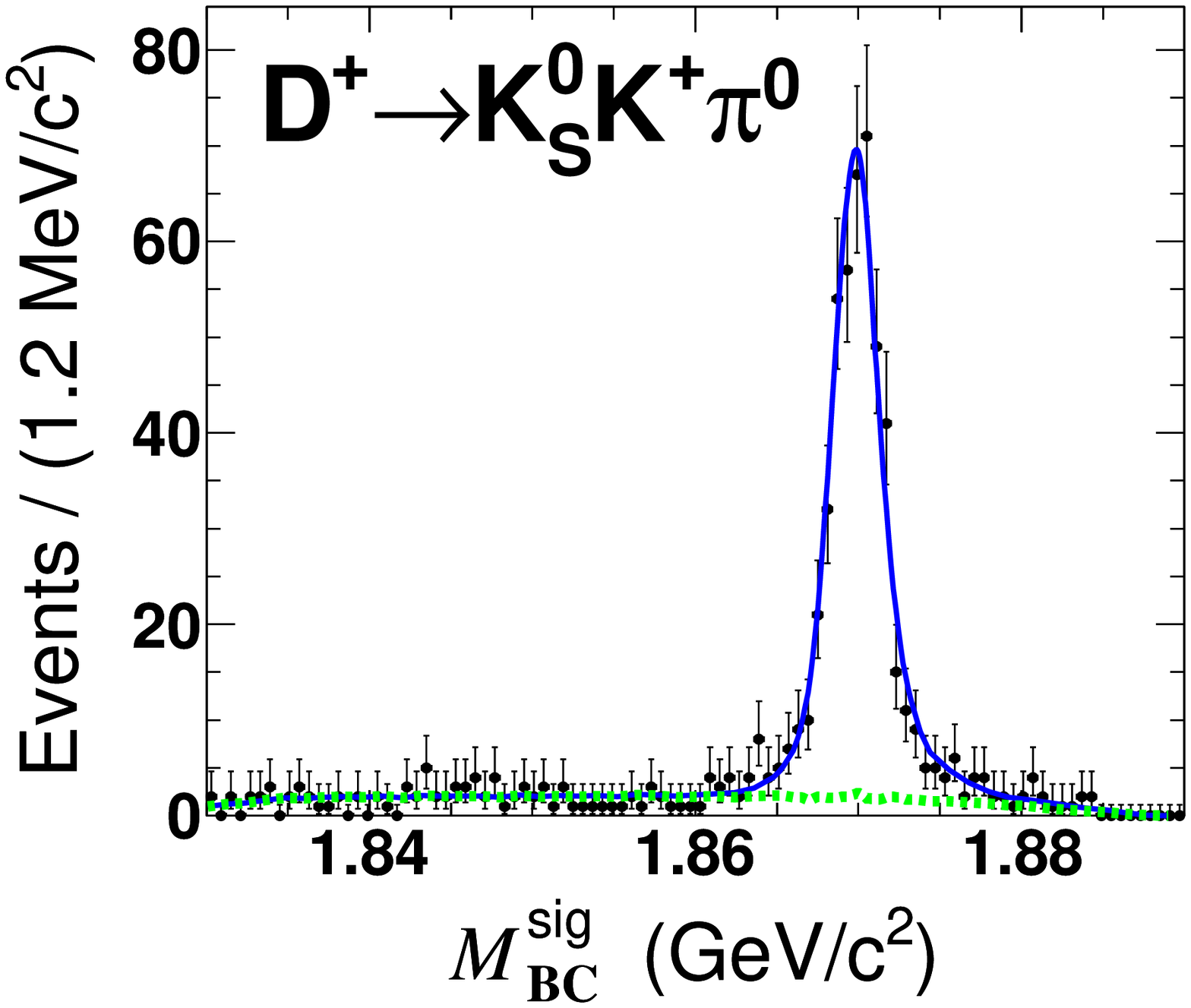}
  \end{overpic}
  \begin{overpic}[width=0.23\linewidth]{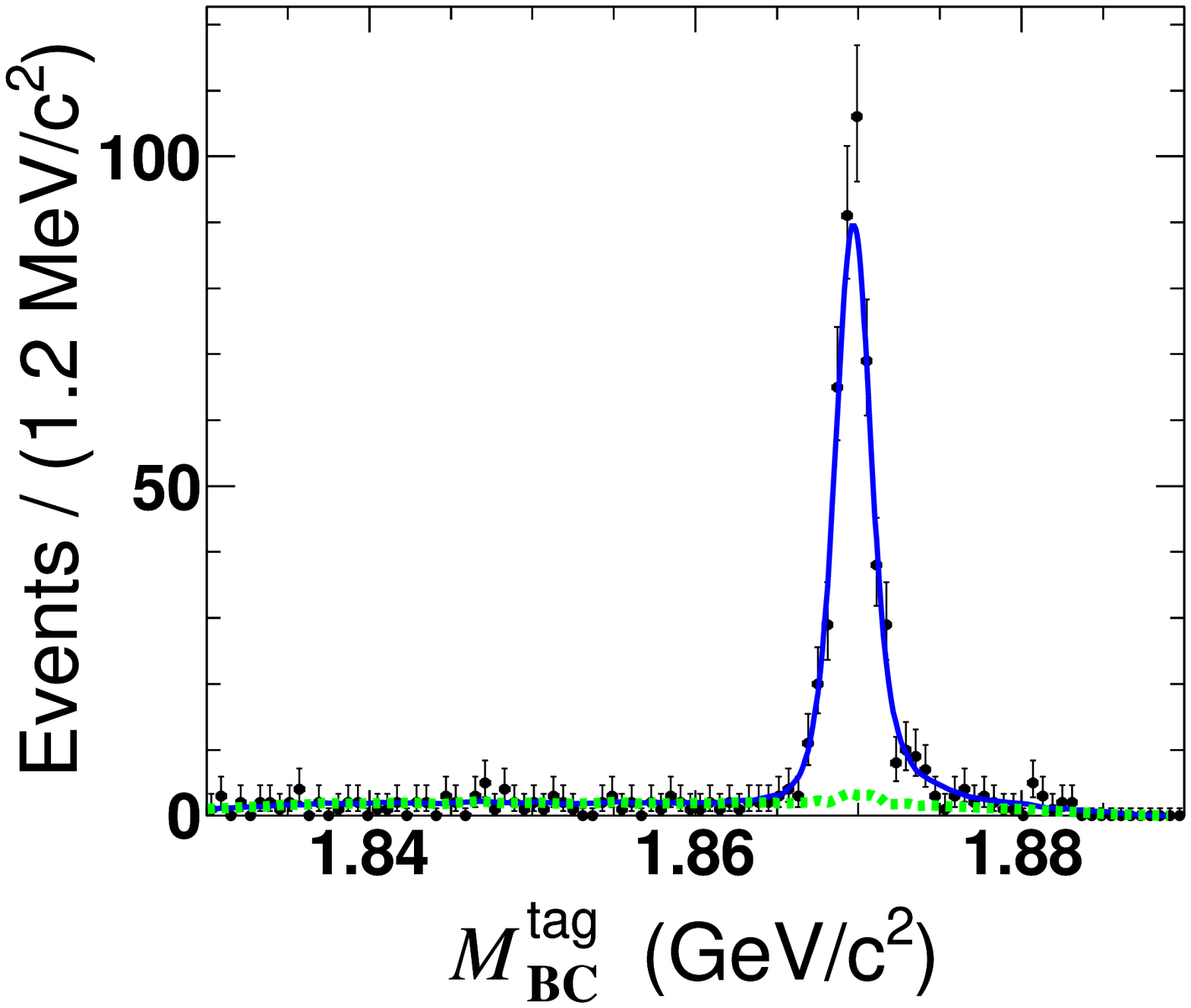}
  \end{overpic}
  \begin{overpic}[width=0.23\linewidth]{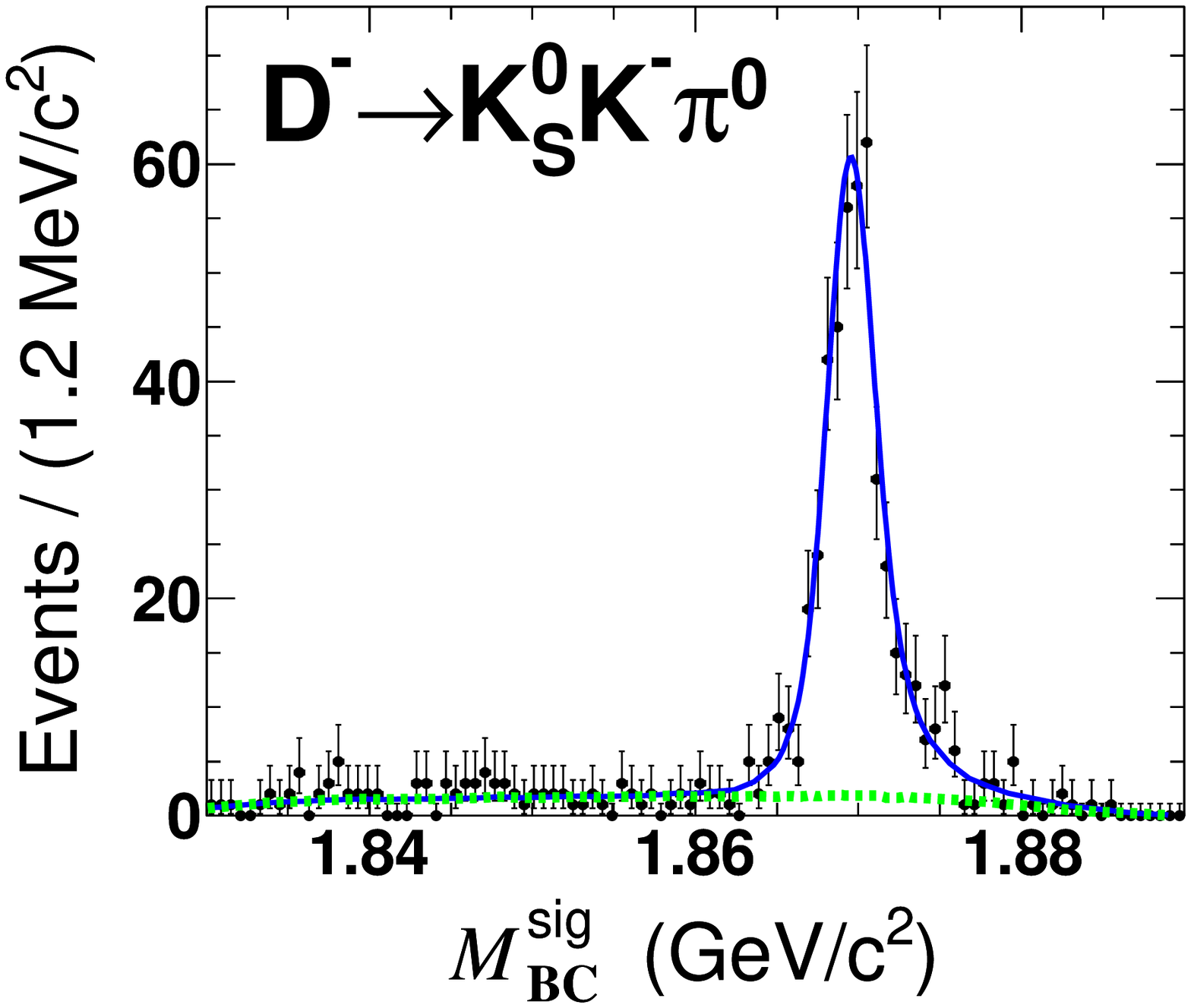}
  \end{overpic}
  \begin{overpic}[width=0.23\linewidth]{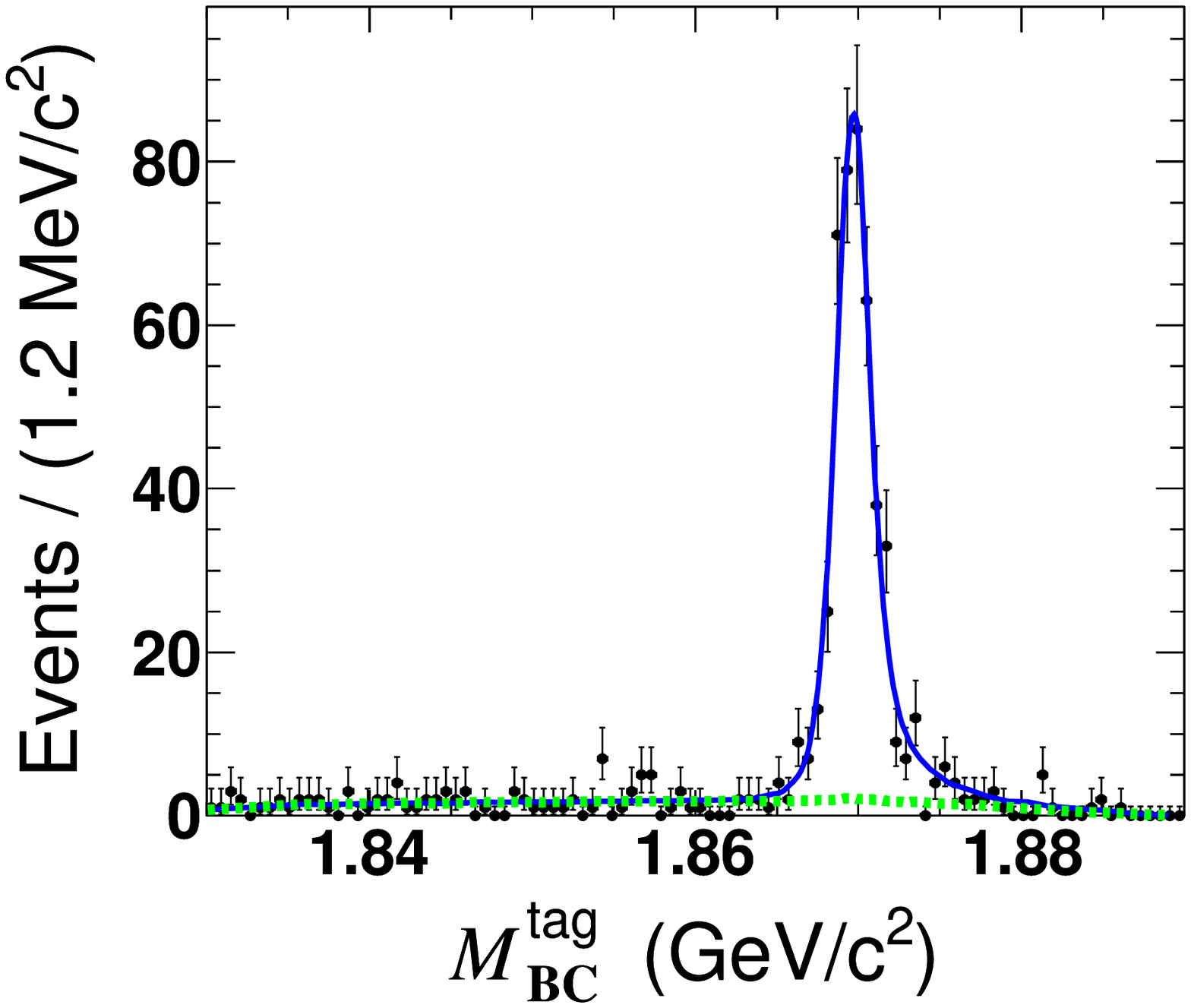}
  \end{overpic}

  \begin{overpic}[width=0.23\linewidth]{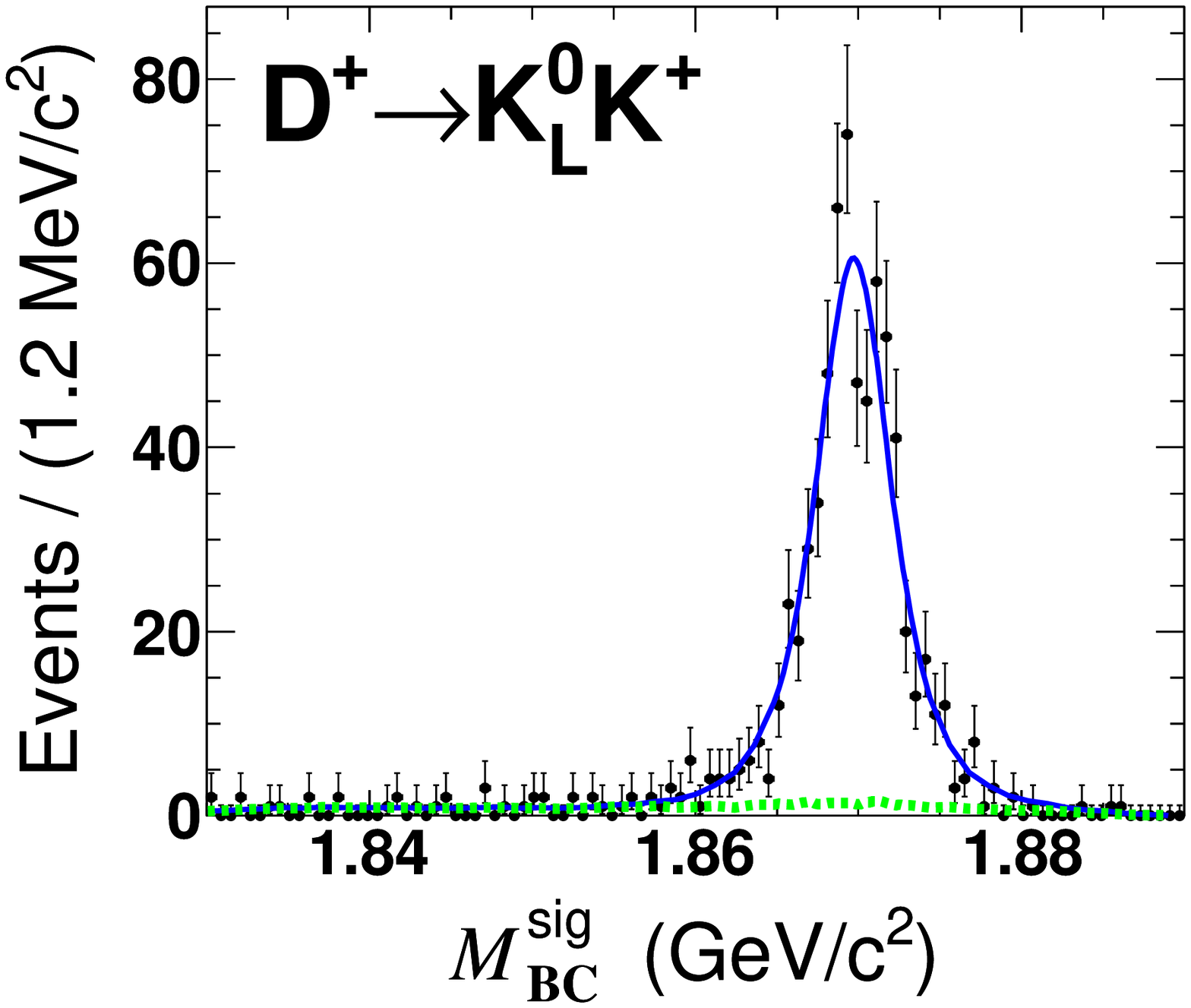}
  \end{overpic}
  \begin{overpic}[width=0.23\linewidth]{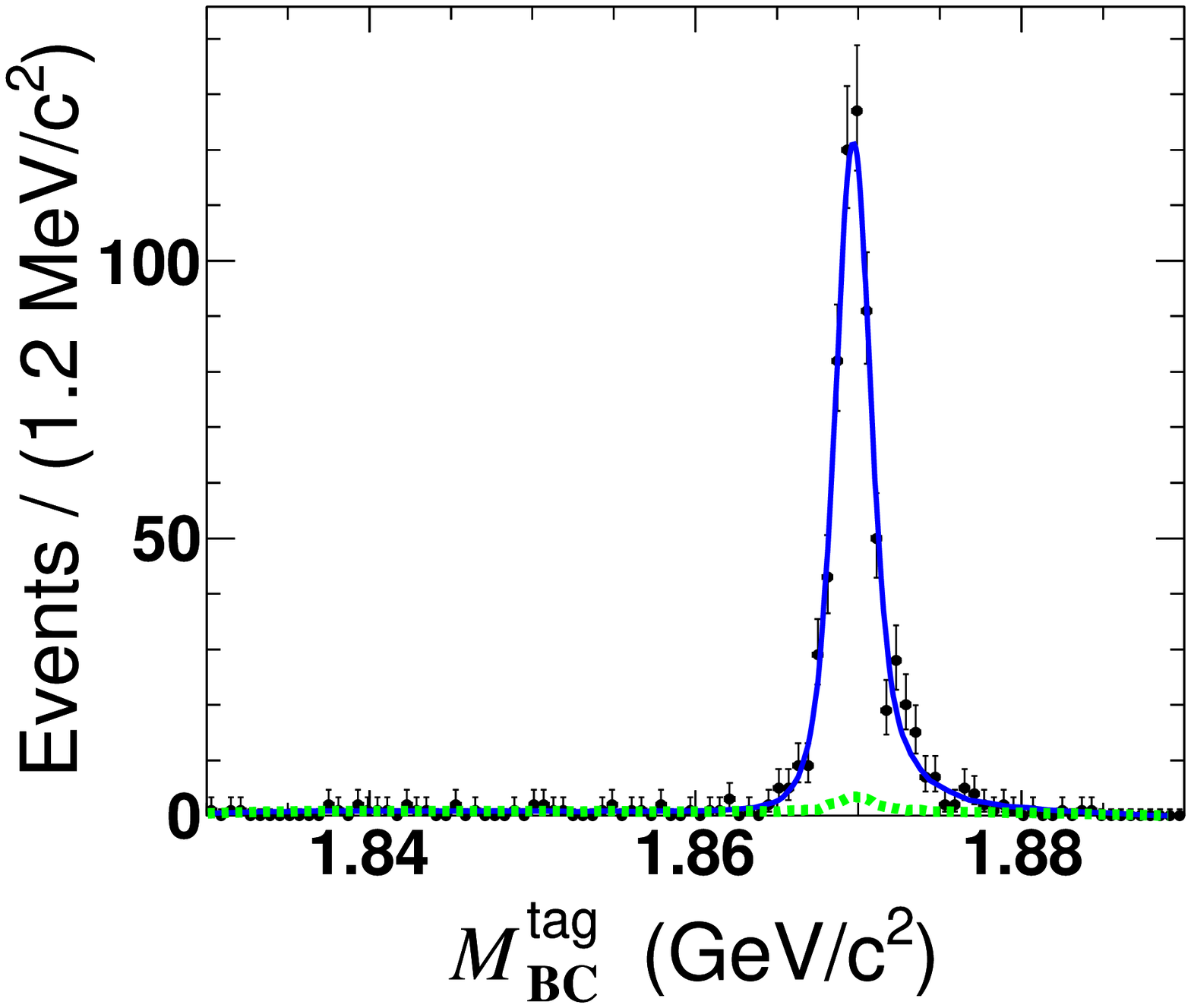}
  \end{overpic}
  \begin{overpic}[width=0.23\linewidth]{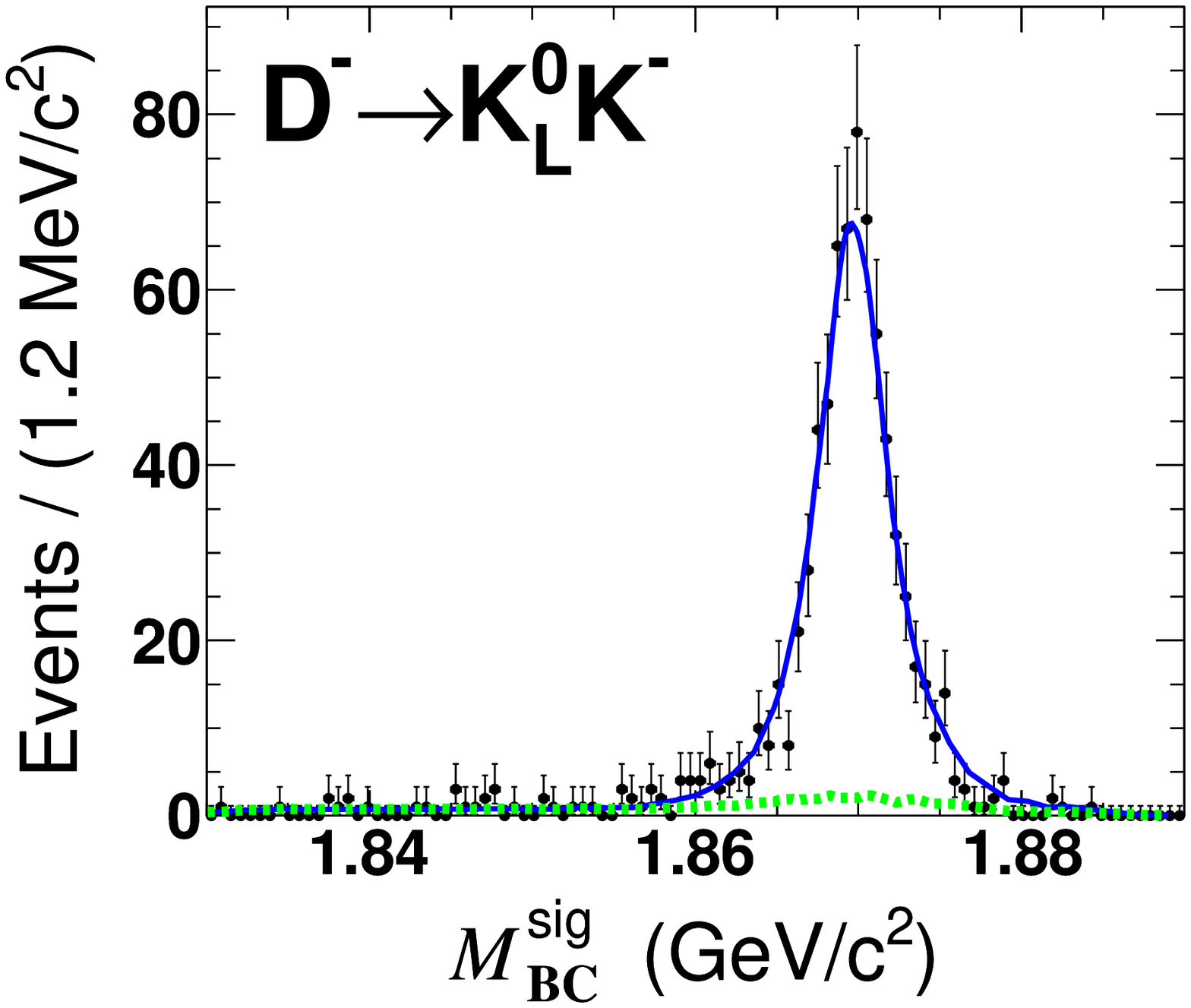}
  \end{overpic}
  \begin{overpic}[width=0.23\linewidth]{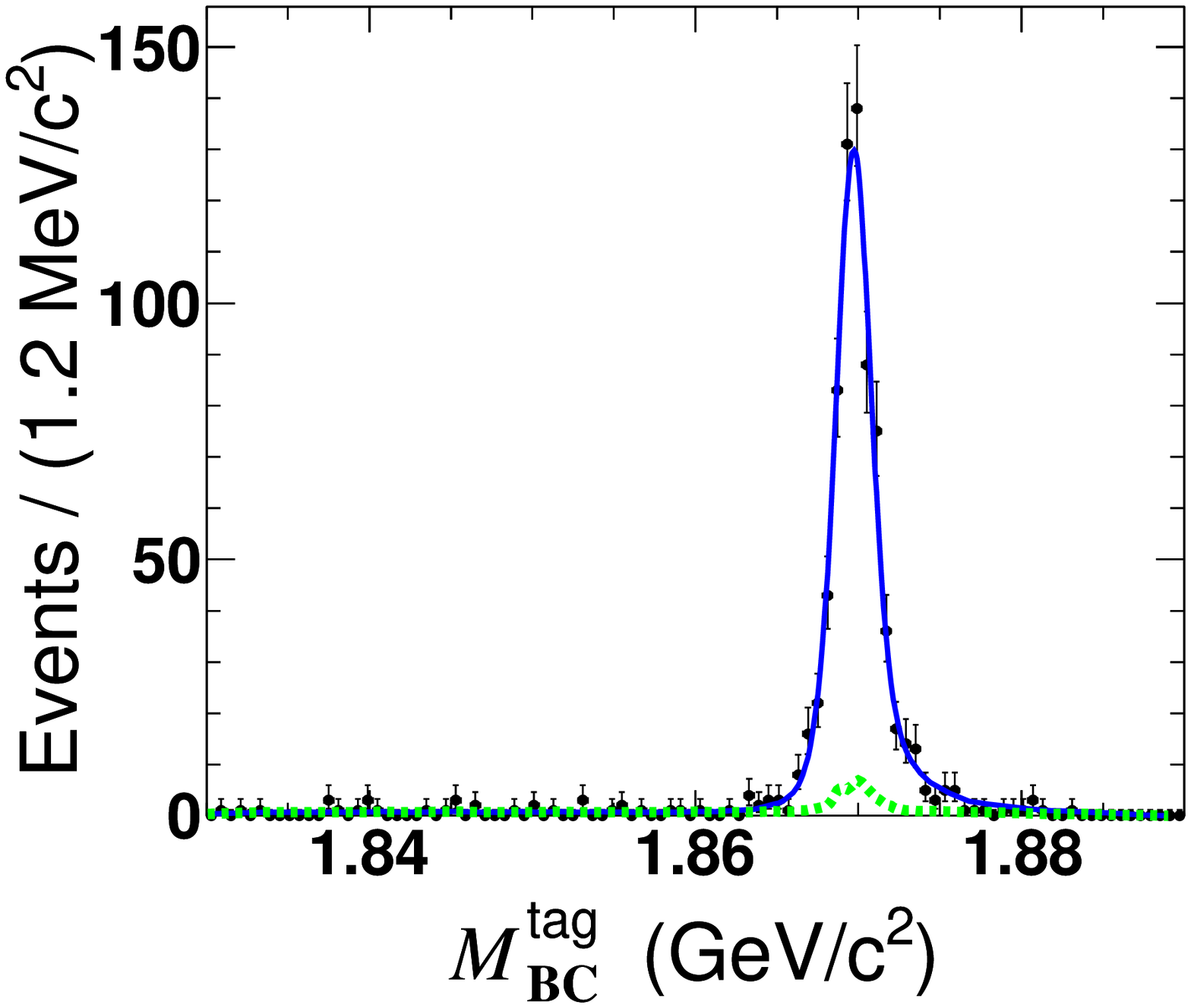}
  \end{overpic}
  \begin{overpic}[width=0.23\linewidth]{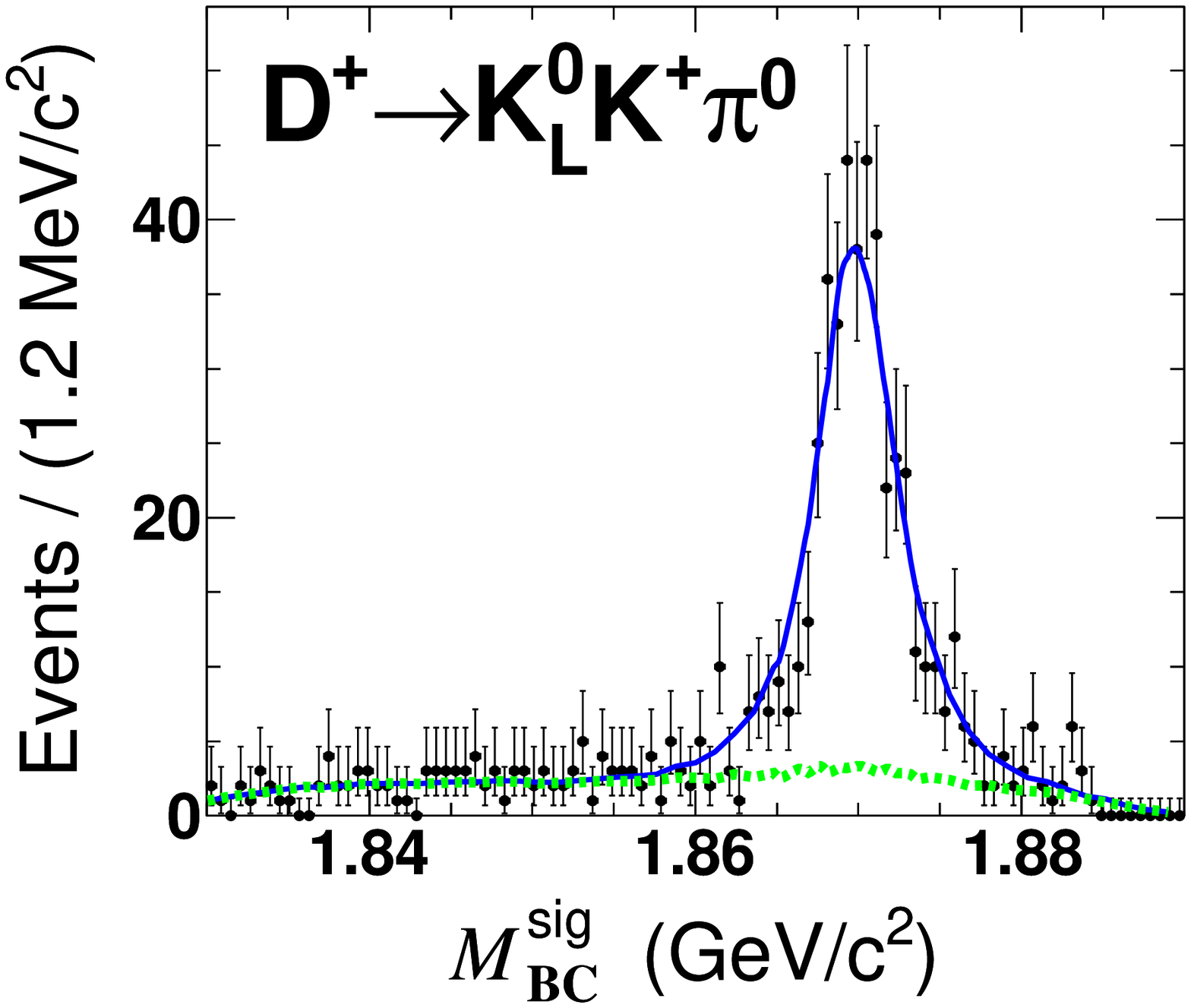}
  \end{overpic}
  \begin{overpic}[width=0.23\linewidth]{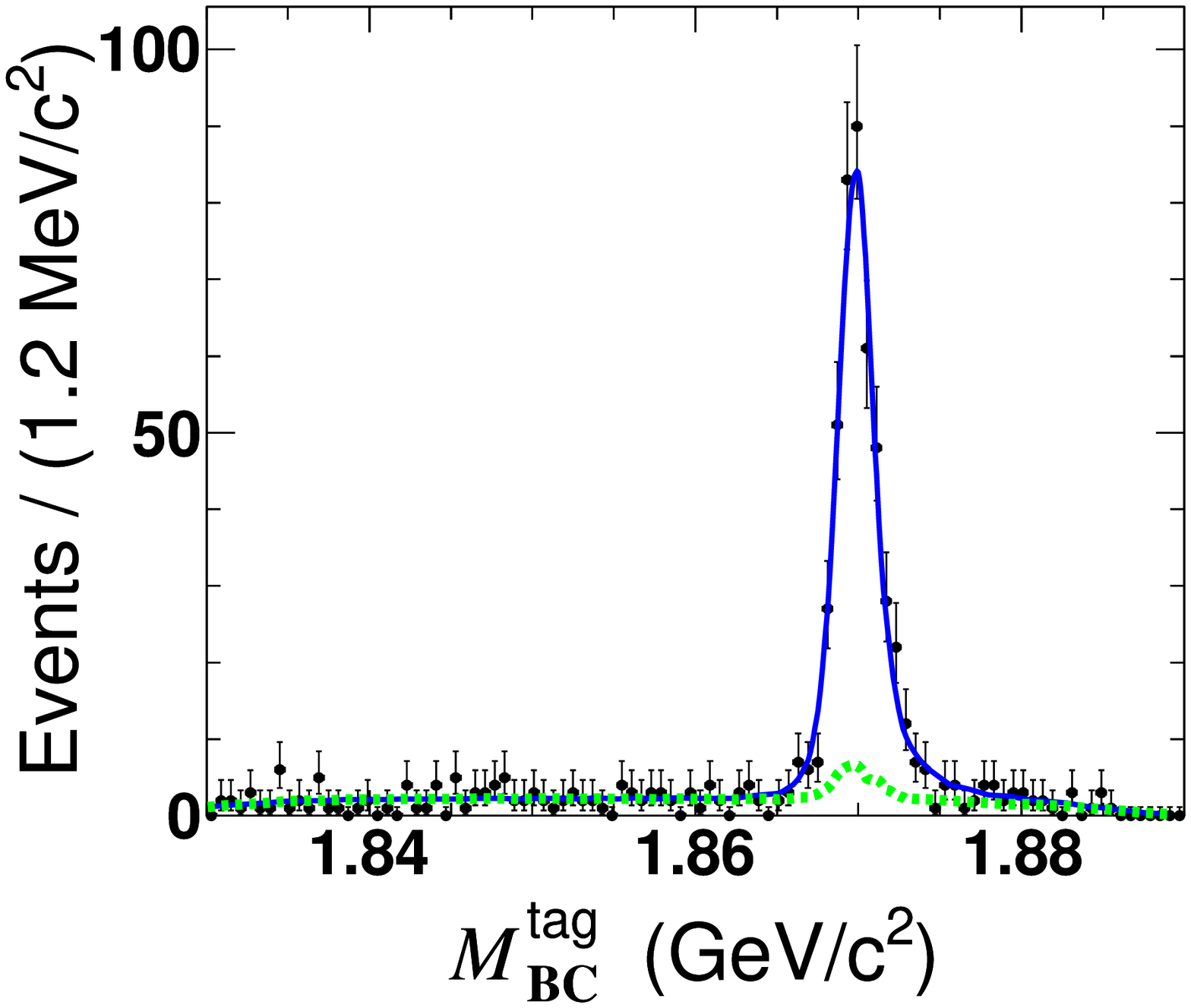}
  \end{overpic}
  \begin{overpic}[width=0.23\linewidth]{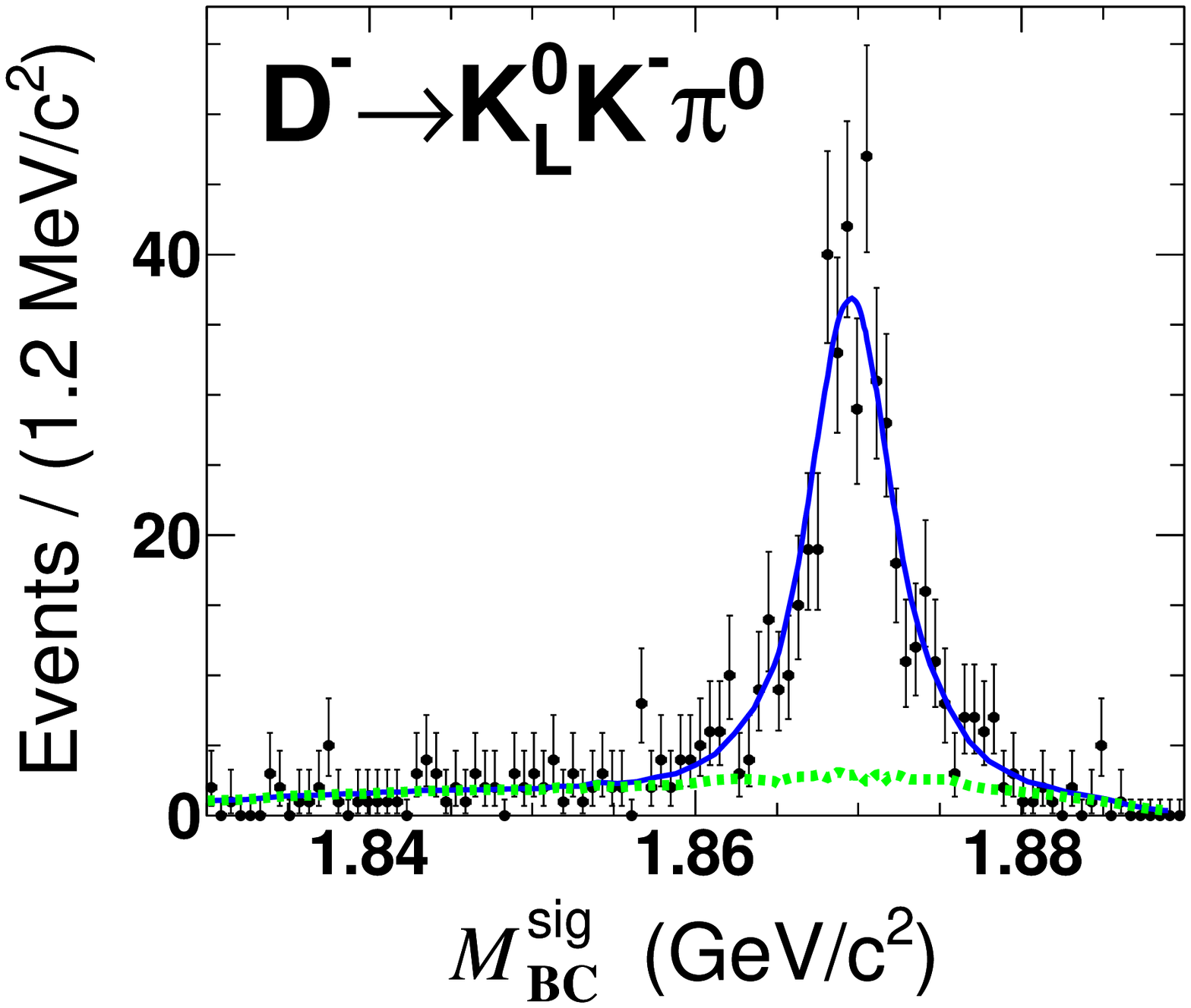}
  \end{overpic}
  \begin{overpic}[width=0.23\linewidth]{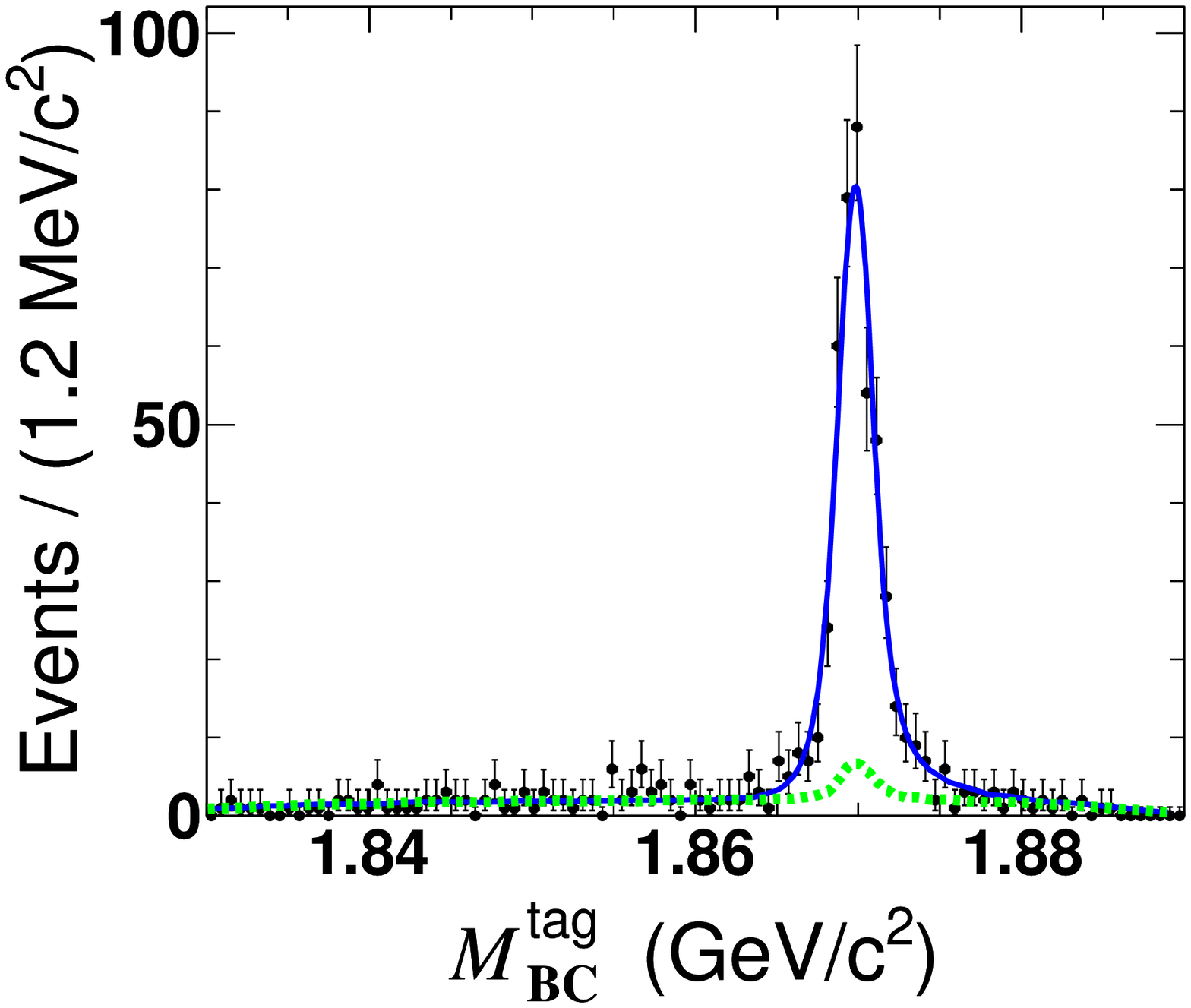}
  \end{overpic}
  \caption{(Color online) Projections on the variables $M_{\rm BC}^{\rm sig}$ and $M_{\rm BC}^{\rm tag}$ of the 2D fits to the signal candidate, summed over all six ST modes. Data are shown as the dots with error bars, the green dashed lines are the backgrounds determined by the fit, and the blue curves are the total fit results.}
  \label{DTfit}
\end{figure*}

The 2D fit is performed on the $M_{\rm BC}^{\rm sig}$ versus $M_{\rm BC}^{\rm tag}$ distribution for each ST mode individually. Figure~\ref{DTfit} shows the projections on the $M_{\rm BC}^{\rm sig}$ and $M_{\rm BC}^{\rm tag}$ distributions of the 2D fits summed over all six ST modes. The detection efficiencies of $D^{\pm} \rightarrow K_{S,L}^0 K^{\pm}(\pi^0)$ are determined by MC simulation. In our previous work \cite{klenu}, differences of the $K_{S,L}^0$ reconstruction efficiencies between data and MC simulation (called data-MC difference) were found, due to differences in nuclear interactions of $K^0$ and $\bar{K^0}$ mesons. The detection efficiencies were investigated for $K^0 \to K_{S,L}^0$ and $\bar{K^0} \to K_{S,L}^0$ separately. To compensate for these differences, the signal efficiencies are corrected by the $K_{S,L}^0$ momentum-weighted data-MC differences of the $K^0_{S,L}$ reconstruction efficiencies. The efficiency correction factors are about 2\% and 10\% for $D^{\pm} \rightarrow K_S^0 K^{\pm}(\pi^0)$ and $D^{\pm} \rightarrow K_L^0 K^{\pm}(\pi^0)$, respectively. The DT signal yields in data ($N_{\rm DT}$) and the corrected detection efficiencies ($\epsilon$) of $D^{\pm} \rightarrow K_{S,L}^0 K^{\pm}(\pi^0)$  are presented in Table~\ref{resultI}.

\begin{table*}[hbtp]
  \centering
  \footnotesize
  \caption{DT yields in data ($N_{\rm DT}$) and efficiencies ($\epsilon$) of reconstructing the signal decays, where the uncertainties are statistical only. The efficiencies include the branching fractions for $K_S^0\rightarrow \pi^+\pi^-$ and $\pi^0 \rightarrow \gamma\gamma$.}\label{resultI}
  \begin{spacing}{1.10}
 \begin{tabular}{L{0.15\linewidth}C{0.13\linewidth}C{0.16\linewidth}|L{0.15\linewidth}
 C{0.13\linewidth}C{0.16\linewidth}}
 \hline\hline
 ST mode & $N_{\rm DT}$ & $\epsilon$ ($\%$) & ST mode & $N_{\rm DT}$ & $\epsilon$ ($\%$) \\
 \hline
 \multicolumn{3}{c|}{$D^-\rightarrow K_S^0 K^-$}   & \multicolumn{3}{c}{$D^+\rightarrow K_S^0 K^+$}\\

$D^{+} \rightarrow K^{-}\pi^{+}\pi^{+}$          & 424 $\pm$ 21 & 34.76   $\pm$  0.43  & $D^{-} \rightarrow K^{+}\pi^{-}\pi^{-}$          & 411 $\pm$ 21  & 34.98    $\pm$ 0.43    \\
$D^{+} \rightarrow K^{-}\pi^{+}\pi^{+}\pi^0$     & 122 $\pm$ 12 & 34.89   $\pm$  0.79  & $D^{-} \rightarrow K^{+}\pi^{-}\pi^{-}\pi^0$     & 133 $\pm$ 11  & 35.24    $\pm$ 0.80    \\
$D^{+} \rightarrow K_S^0\pi^{+}$                 & 68  $\pm$ 9  & 34.27   $\pm$  1.30  & $D^{-} \rightarrow K_S^0\pi^{-}$                 & 41  $\pm$ 7   & 34.34    $\pm$ 1.30    \\
$D^{+} \rightarrow K_S^0\pi^{+}\pi^0$            & 114 $\pm$ 11 & 34.28   $\pm$  0.87  & $D^{-} \rightarrow K_S^0\pi^{-}\pi^0$            & 112 $\pm$ 11  & 33.82    $\pm$ 0.87    \\
$D^{+} \rightarrow K_S^0\pi^{+}\pi^{+}\pi^{-}$   & 57  $\pm$ 8  & 33.30   $\pm$  1.10  & $D^{-} \rightarrow K_S^0\pi^{-}\pi^{-}\pi^{+}$   & 60  $\pm$ 9   & 32.32    $\pm$ 1.10    \\
$D^{+} \rightarrow K^{-}K^{+}\pi^{+}$            & 37  $\pm$ 7  & 35.27   $\pm$  1.50  & $D^{-} \rightarrow K^{+}K^{-}\pi^{-}$            & 39  $\pm$ 7   & 36.20    $\pm$ 1.50    \\  \hline

 \multicolumn{3}{c|}{{\color{black}{$D^-\rightarrow K_S^0 K^-\pi^0$}}}   &  \multicolumn{3}{c}{{\color{black}{$D^+\rightarrow K_S^0 K^+\pi^0$}}}\\

$D^{+} \rightarrow K^{-}\pi^{+}\pi^{+}$          &248  $\pm$  16  & 12.00  $\pm$  0.20  &  $D^{-} \rightarrow K^{+}\pi^{-}\pi^{-}$          &253  $\pm$  17  & 12.06  $\pm$  0.20\\
$D^{+} \rightarrow K^{-}\pi^{+}\pi^{+}\pi^0$     &65   $\pm$  9   & 10.64  $\pm$  0.37  &  $D^{-} \rightarrow K^{+}\pi^{-}\pi^{-}\pi^0$     &71   $\pm$  9   & 11.18  $\pm$  0.37 \\
$D^{+} \rightarrow K_S^0\pi^{+}$                 &23   $\pm$  5   & 11.85  $\pm$  0.59  &  $D^{-} \rightarrow K_S^0\pi^{-}$                 &25   $\pm$  6   & 11.98  $\pm$  0.58 \\
$D^{+} \rightarrow K_S^0\pi^{+}\pi^0$            &60   $\pm$  8   & 11.26  $\pm$  0.40  &  $D^{-} \rightarrow K_S^0\pi^{-}\pi^0$            &63   $\pm$  9   & 12.04  $\pm$  0.42 \\
$D^{+} \rightarrow K_S^0\pi^{+}\pi^{+}\pi^{-}$   &29   $\pm$  6   & 10.19  $\pm$  0.49  &  $D^{-} \rightarrow K_S^0\pi^{-}\pi^{-}\pi^{+}$   &35   $\pm$  7   & 10.76  $\pm$  0.49 \\
$D^{+} \rightarrow K^{-}K^{+}\pi^{+}$            &19   $\pm$  6   & 11.15  $\pm$  0.64  &  $D^{-} \rightarrow K^{+}K^{-}\pi^{-}$            &22   $\pm$  6   & 11.31  $\pm$  0.67 \\ \hline

 \multicolumn{3}{c|}{{\color{black}{$D^-\rightarrow K_L^0 K^-$}}}   & \multicolumn{3}{c}{{\color{black}{$D^+\rightarrow K_L^0 K^+$}}}\\
$D^{+} \rightarrow K^{-}\pi^{+}\pi^{+}$          & 375 $\pm$  21 &   27.43 $\pm$  0.39  &   $D^{-} \rightarrow K^{+}\pi^{-}\pi^{-}$          & 343 $\pm$  19 &   27.96 $\pm$  0.39 \\
$D^{+} \rightarrow K^{-}\pi^{+}\pi^{+}\pi^0$     & 94  $\pm$  10 &   24.24 $\pm$  0.69  &   $D^{-} \rightarrow K^{+}\pi^{-}\pi^{-}\pi^0$     & 92  $\pm$  10 &   26.50 $\pm$  0.70 \\
$D^{+} \rightarrow K_S^0\pi^{+}$                 & 40  $\pm$  7  &   27.61 $\pm$  1.20  &   $D^{-} \rightarrow K_S^0\pi^{-}$                 & 41  $\pm$  7  &   28.99 $\pm$  1.20 \\
$D^{+} \rightarrow K_S^0\pi^{+}\pi^0$            & 89  $\pm$  10 &   25.19 $\pm$  0.77  &   $D^{-} \rightarrow K_S^0\pi^{-}\pi^0$            & 105 $\pm$  11 &   27.93 $\pm$  0.78 \\
$D^{+} \rightarrow K_S^0\pi^{+}\pi^{+}\pi^{-}$   & 41  $\pm$  7  &   21.87 $\pm$  0.99  &   $D^{-} \rightarrow K_S^0\pi^{-}\pi^{-}\pi^{+}$   & 44  $\pm$  7  &   21.98 $\pm$  0.97 \\
$D^{+} \rightarrow K^{-}K^{+}\pi^{+}$            & 31  $\pm$  6  &   23.95 $\pm$  1.30  &   $D^{-} \rightarrow K^{+}K^{-}\pi^{-}$            & 23  $\pm$  6  &   21.79 $\pm$  1.20 \\ \hline

 \multicolumn{3}{c|}{{\color{black}{$D^-\rightarrow K_L^0 K^-\pi^0$}}}   & \multicolumn{3}{c}{{\color{black}{$D^+\rightarrow K_L^0 K^+\pi^0$}}}\\

$D^{+} \rightarrow K^{-}\pi^{+}\pi^{+}$         & 250 $\pm$  17 &   11.01 $\pm$  0.18  &  $D^{-} \rightarrow K^{+}\pi^{-}\pi^{-}$        & 241 $\pm$  17 &   11.31 $\pm$  0.18  \\
$D^{+} \rightarrow K^{-}\pi^{+}\pi^{+}\pi^0$    & 48  $\pm$  8  &   9.20  $\pm$  0.32  &  $D^{-} \rightarrow K^{+}\pi^{-}\pi^{-}\pi^0$   & 65  $\pm$  9  &   9.17  $\pm$  0.32  \\
$D^{+} \rightarrow K_S^0\pi^{+}$                & 23  $\pm$  5  &   10.20 $\pm$  0.54  &  $D^{-} \rightarrow K_S^0\pi^{-}$               & 25  $\pm$  6  &   10.71 $\pm$  0.55  \\
$D^{+} \rightarrow K_S^0\pi^{+}\pi^0$           & 58  $\pm$  9  &   8.93  $\pm$  0.34  &  $D^{-} \rightarrow K_S^0\pi^{-}\pi^0$          & 48  $\pm$  8  &   9.53  $\pm$  0.35  \\
$D^{+} \rightarrow K_S^0\pi^{+}\pi^{+}\pi^{-}$  & 19  $\pm$  5  &   7.94  $\pm$  0.44  &  $D^{-} \rightarrow K_S^0\pi^{-}\pi^{-}\pi^{+}$ & 23  $\pm$  6  &   7.55  $\pm$  0.42  \\
$D^{+} \rightarrow K^{-}K^{+}\pi^{+}$           & 14  $\pm$  5  &   8.03  $\pm$  0.55  &  $D^{-} \rightarrow K^{+}K^{-}\pi^{-}$          & 14  $\pm$  5  &   8.71  $\pm$  0.57  \\

\hline
\hline
 \end{tabular}
 \end{spacing}
\end{table*}

\subsection{\boldmath Branching fraction and $CP$ asymmetry}

According to Eq.~(\ref{BF}) and taking into account the numbers of $N_{\rm ST}$, $N_{\rm DT}$, and $\epsilon$ listed in Tables~\ref{ST} and \ref{resultI}, the branching fractions of $D^+$ and $D^-$ decays for the individual ST modes are calculated. The average branching fractions of $D^+$ and $D^-$ decays as well as combination of charged conjugation modes are obtained by using the standard weighted least-squares method~\cite{PDG}, and are summarized in Table~\ref{resultII}. We also determine the $CP$ asymmetries with Eq.~(\ref{acp}) based on the average branching fractions of $D^+$ and $D^-$ decays, and the results are listed in Table~\ref{resultII}, too.

\begin{table*}[hbtp]
  \centering
  \footnotesize
  \caption{The measured branching fractions and $CP$ asymmetries, where the first and second uncertainties are statistical and systematic, respectively, and a comparison with the world average value~\cite{PDG}.}\label{resultII}
  \begin{spacing}{1.29}
   \begin{tabular}{L{0.1\linewidth}C{0.15\linewidth}C{0.15\linewidth}C{0.15\linewidth}
   C{0.15\linewidth}C{0.15\linewidth}}
   \hline\hline
  Signal mode            &  $\mathcal{B}(D^+)$ ($\times10^{-3}$)   &   $\mathcal{B}(D^-)$ ($\times10^{-3}$)  &  $\mathcal{\overline B}$ ($\times10^{-3}$)
  &  $\mathcal B$ (PDG) ($\times10^{-3}$)  &  $\mathcal{A}_{CP}$ ($\%$)  \\
 \hline

 $K_S^0 K^{\pm}$       &  2.96  $\pm$  0.11  $\pm$ 0.08    &   3.07  $\pm$  0.12  $\pm$ 0.08    &  3.02  $\pm$  0.09  $\pm$ 0.08 & 2.95 $\pm$ 0.15 & -1.8 $\pm$ 2.7 $\pm$ 1.6\\
 $K_S^0 K^{\pm}\pi^0$  &  5.14  $\pm$  0.27  $\pm$ 0.24    &   5.00  $\pm$  0.26  $\pm$ 0.22    &  5.07  $\pm$  0.19  $\pm$ 0.23 & - & 1.4  $\pm$ 3.7 $\pm$ 2.4 \\
 $K_L^0 K^{\pm}$       &  3.07  $\pm$  0.14  $\pm$ 0.10    &   3.34  $\pm$  0.15  $\pm$ 0.11    &  3.21  $\pm$  0.11  $\pm$ 0.11 & - & -4.2 $\pm$ 3.2 $\pm$ 1.2 \\
 $K_L^0 K^{\pm}\pi^0$  &  5.21  $\pm$  0.30  $\pm$ 0.22    &   5.27  $\pm$  0.30  $\pm$ 0.22    &  5.24  $\pm$  0.22  $\pm$ 0.22 & - & -0.6 $\pm$ 4.1 $\pm$ 1.7 \\
\hline
\hline
 \end{tabular}
 \end{spacing}
\end{table*}

\section{SYSTEMATIC UNCERTAINTY}\label{sysBr}

Due to the use of the DT method, those uncertainties associated with the ST selection are cancelled.
The relative systematic uncertainties in the measurements of absolute branching fractions and the $CP$ asymmetries of the decay $D^\pm\rightarrow K_{S,L}^0 K^\pm(\pi^0)$ are summarized in Table~\ref{mysys} and are discussed in detail below.

\begin{table*}[hbtp]
  \centering
  \footnotesize
  \caption{Systematic uncertainties ($\%$) of the measured branching fractions and corresponding $CP$ asymmetries.}\label{mysys}
  \begin{spacing}{1.29}
  \begin{tabular}{L{0.2\linewidth}C{0.09\linewidth}C{0.09\linewidth}C{0.09\linewidth}
   C{0.09\linewidth}C{0.09\linewidth}C{0.09\linewidth}C{0.09\linewidth}C{0.09\linewidth}}
 \hline\hline
 Source & $K_S^0 K^+$ & $K_S^0 K^-$  &  $K_S^0K^+\pi^0$ & $K_S^0K^-\pi^0$
                   & $K_L^0 K^+$ & $K_L^0 K^-$  &  $K_L^0K^+\pi^0$ & $K_L^0K^-\pi^0$  \\
 \hline
 $K^{\pm}$ tracking              &0.7        & 0.9      &1.8        & 1.4     &0.7        & 0.8     &1.8        & 1.5          \\
 $K^{\pm}$ PID                   &0.3        & 0.2      &0.2        & 0.3     &0.3        & 0.2     &0.2        & 0.3          \\
 $\pi^0$ reconstruction          & -         & -        &2.0        & 2.0     & -         & -       & 2.0       & 2.0          \\
 $K_S^0$ reconstruction          &1.9        & 1.9      &2.9        & 2.8     & -         & -       & -         & -            \\
 $K_L^0$ reconstruction          & -         & -        & -         & -       &1.2        & 1.3     &1.4        & 1.4          \\
 $\chi^2_{K^0_L}$ cut            & -         & -        & -         & -       &2.5        & 2.5     &1.7        & 1.8          \\
 Sub-resonances                  & -         & -        &1.4        & 1.1     & -         & -       &1.5        & 1.5          \\
 $M_{\rm BC}$ fit in DT          & 1.3       & 1.3      & 1.5       & 1.5     & 1.1       & 1.1     &  1.6      &  1.6         \\
 Peaking backgrounds in DT       & -         &  -       & -         & -       &0.1        & 0.1     &0.2        & 0.2          \\
 $\Delta E$ requirement          & 0.6       & 0.6      & 0.6       &0.6      &-          &-        & -         & -            \\
 \hline
 \quad Total (for $\mathcal{B}$)                  & 2.5       & 2.6      & 4.5       & 4.2     & 3.1       & 3.2     &  4.2      & 4.1       \\
 \quad Total (for $\mathcal{A}_{CP}$)             & 2.1       & 2.2      & 3.5       & 3.2     & 1.5       & 1.6     &  2.3      & 2.1       \\
\hline
\hline
 \end{tabular}
 \end{spacing}
\end{table*}

The efficiencies of $K^{\pm}$ tracking and PID in various $K^{\pm}$ momentum ranges are investigated with $K^{\pm}$ samples selected from DT hadronic $D\bar D$ decays. In each momentum range, the data-MC difference of efficiencies $\epsilon_{\rm data}/\epsilon_{\rm MC}-1$ is calculated. The data-MC differences weighted by the $K^\pm$ momentum in the decays $D^\pm\rightarrow K_{S,L}^0 K^\pm(\pi^0)$ are assigned as the associated systematic uncertainties.

The $\pi^0$ reconstruction efficiency is studied by the DT control sample $D^0 \rightarrow K^-\pi^+\pi^0$ versus $\bar D^0 \rightarrow K^+\pi^-$ or $\bar D^0 \rightarrow K^+\pi^-\pi^-\pi^+$ using the partial reconstruction technique. The data-MC difference of the $\pi^0$ reconstruction efficiencies weighted according to the $\pi^0$ momentum distribution in $D^+\rightarrow K_{S,L}^0 K^+ \pi^0$ is assigned as the systematic uncertainty in $\pi^0$ reconstruction.

The branching fractions of $K^0_S \to \pi^+\pi^-$ and $\pi^0 \to \gamma\gamma$ are taken from the Particle Data Group~\cite{PDG}.
Their uncertainties are $0.07\%$ and $0.03\%$, respectively, which are negligible in these measurements.

As described in Ref.~\cite{klenu}, the correction factors of $K_{S,L}^0$ reconstruction efficiencies are determined with the two control samples of $J/\psi \rightarrow K^\ast(892)^\mp K^\pm$ with $K^\ast(892)^\mp \rightarrow K_{S,L}^0\pi^\mp$ and $J/\psi \rightarrow \phi K_{S,L}^0 K^\pm \pi^\mp$ decays.  Since the efficiency corrections are imposed in this analysis, the corresponding statistical uncertainties of the correction factors, which are weighted according to the $K^0_{S,L}$ momentum in the decays $D^\pm \to K^0_{S,L} K^\pm(\pi^0)$, are assigned as the uncertainty associated with the $K^0_{S,L}$ reconstruction efficiency.

As described in Ref.~\cite{klenu}, in the determination of the correction factor of the $K_L^0$ efficiency, we perform a kinematic fit to select the $K_L^0$ candidate with the minimal $\chi_{K_L^0}^2$ and require $\chi_{K_L^0}^2 <100$. The uncertainty of the correction factor associated with the $\chi_{K_L^0}^2$ cut is determined by comparing the selection efficiency between data and MC simulation using the same control samples. The $\chi_{K_L^0}^2$ requirement summarized in Table~\ref{Tchisq} brings an uncertainty. The momentum-weighted uncertainty of the $\chi_{K_L^0}^2$ selection according to the $K_L^0$ momentum distribution of signal events is assigned as the associated systematic uncertainties.

In the analysis of multi-body decays, the detection efficiency may depend on the kinematic variables of the final-state particles. The possible difference of the kinematic variable distribution between data and MC simulation causes an uncertainty on detection efficiency. For the three-body decays  $D^+ \rightarrow K_{S,L}^0 K^+ \pi^0$, the nominal efficiencies are estimated by analyzing an MC sample composed of the decays $D^+\to K^{\ast}(892)^+ \bar K^0$, $D^+\to \bar K^{\ast}(892)^0 K^+$, $D^+\to \bar K^{\ast}(1430)^0 K^+$, and $D^+\to \bar K_2^{\ast}(1430)^0 K^+$. The fractions of these components are taken from the Dalitz plot analysis of the charge conjugated decay $D^+ \rightarrow K^+K^-\pi^+$ \cite{kkpi}. The differences of the nominal efficiencies to those estimated with an MC sample of their dominant decays of $D^+\to K^{\ast}(892)^+ K_{S,L}^0$~\cite{PDG} are taken as the systematic uncertainties due to the MC model.

To evaluate the systematic uncertainty associated with the ST yields, we repeat the fit on the $M_{\rm BC}$ distribution of ST candidate events by varying the resolution of the Gaussian function by one standard deviation. The resulting change on the ST yields is found to be negligible.

The systematic uncertainties in the 2D fit on the $M_{\rm BC}^{\rm tag}$ versus $M_{\rm BC}^{\rm sig}$ distribution are evaluated by repeating the fit with an alternative fit range $(1.8400,1.8865)$ GeV/$c^2$, varying the resolution of the smearing Gaussian function by one standard deviation, and varying the endpoint of the ARGUS function by $\pm0.2$ MeV/$c^2$, individually, and the sum in quadrature of the changes in DT yields are taken as the systematic uncertainties.

As described in Sec.~\ref{klrec}, the dominant peaking backgrounds for $D^\pm\rightarrow K_L^0 K^\pm(\pi^0)$ are found to be from $D^\pm\rightarrow K_S^0 K^\pm(\pi^0)$ with $K_S^0 \rightarrow \pi^0\pi^0$, whose contributions are about $3\%~(5\%)$. Their sizes are estimated based on MC simulation after considering the branching fraction of the background channel and are fixed in the fits. Other peaking backgrounds like $D^\pm\rightarrow K_L^0 \pi^\pm(\pi^0)$ are found to have contributions of less than $0.5\%$. The uncertainties due to these peaking backgrounds are estimated by varying the branching fractions of the peaking background channels by $\pm 1\sigma$, and the changes of the DT signal yields are assigned as the associated systematic uncertainties.

In the studies of $D^\pm\rightarrow K_S^0K^\pm(\pi^0)$, a $\Delta E$ requirement in the signal side is applied to suppress the background. The corresponding uncertainty is studied by comparing the DT yields with and without the $\Delta E$ requirement for an ST mode with low background, $i.e.$  $D^\pm\rightarrow K^\mp\pi^\pm\pi^\pm$. The resulting difference of relative change of DT yields between data and MC simulation is assigned as the systematic uncertainty.

For each signal mode, the total systematic uncertainty of the measured branching fraction is obtained by adding all above individual uncertainties in quadrature, as summarized in Table~\ref{mysys}. In the determination of the $CP$ asymmetries, the uncertainties arising from $\pi^0$ reconstruction, $\chi_{K_L^0}^2$ requirement of the $K_L^0$ selection, MC model of $D^{\pm} \rightarrow K_{S,L}^0 K^{\pm}\pi^0$, $M_{\rm BC}$ fit for ST events and $\Delta E$ requirement in signal side are canceled. The total systematic uncertainties in the measured $CP$ asymmetries are also listed in Table~\ref{mysys}.

\section{\boldmath $CP$ Asymmetries in different Dalitz plot regions for $D^\pm\rightarrow K_{S,L}^0 K^\pm\pi^0$}

We also examine the $CP$ asymmetries for the three-body decay $D^\pm\rightarrow K_{S,L}^0 K^\pm\pi^0$ in different regions across the Dalitz plot. For this study, a further kinematic fit constraining the masses of $K_S^0$ and $D^+$ candidates to their nominal masses~\cite{PDG} is performed in the selection of $D^\pm\rightarrow K_S^0 K^\pm\pi^0$.  To select signal $D^\pm\rightarrow K_L^0 K^\pm\pi^0$ events, a kinematic fit constraining the $D^+$ to its nominal mass is performed in addition to the kinematic fit to select the $K_L^0$ shower as described in Sec.~\ref{klrec}. The recoiling mass of the $K_{S,L}^0 K^\pm\pi^0$ system,

\begin{equation}
M_{\rm rec} = \sqrt{(q_0 - q_D)^2},
\end{equation}\label{cpasy}

\noindent which should equal the mass of the ST $D$ meson in correctly reconstructed signal events, is used to identify the signal, where $q_0$ and $q_D$ are the four-momentum of the $e^+e^-$ system and the selected $D^+$ candidate, respectively. This procedure ensures that $D$ candidates have the same phase space~$(\rm PHSP)$, regardless of whether $M_{\rm rec}$ is in the signal or sideband region.

\begin{figure}[hbtp]
  \centering
  \includegraphics[width=0.49\linewidth]{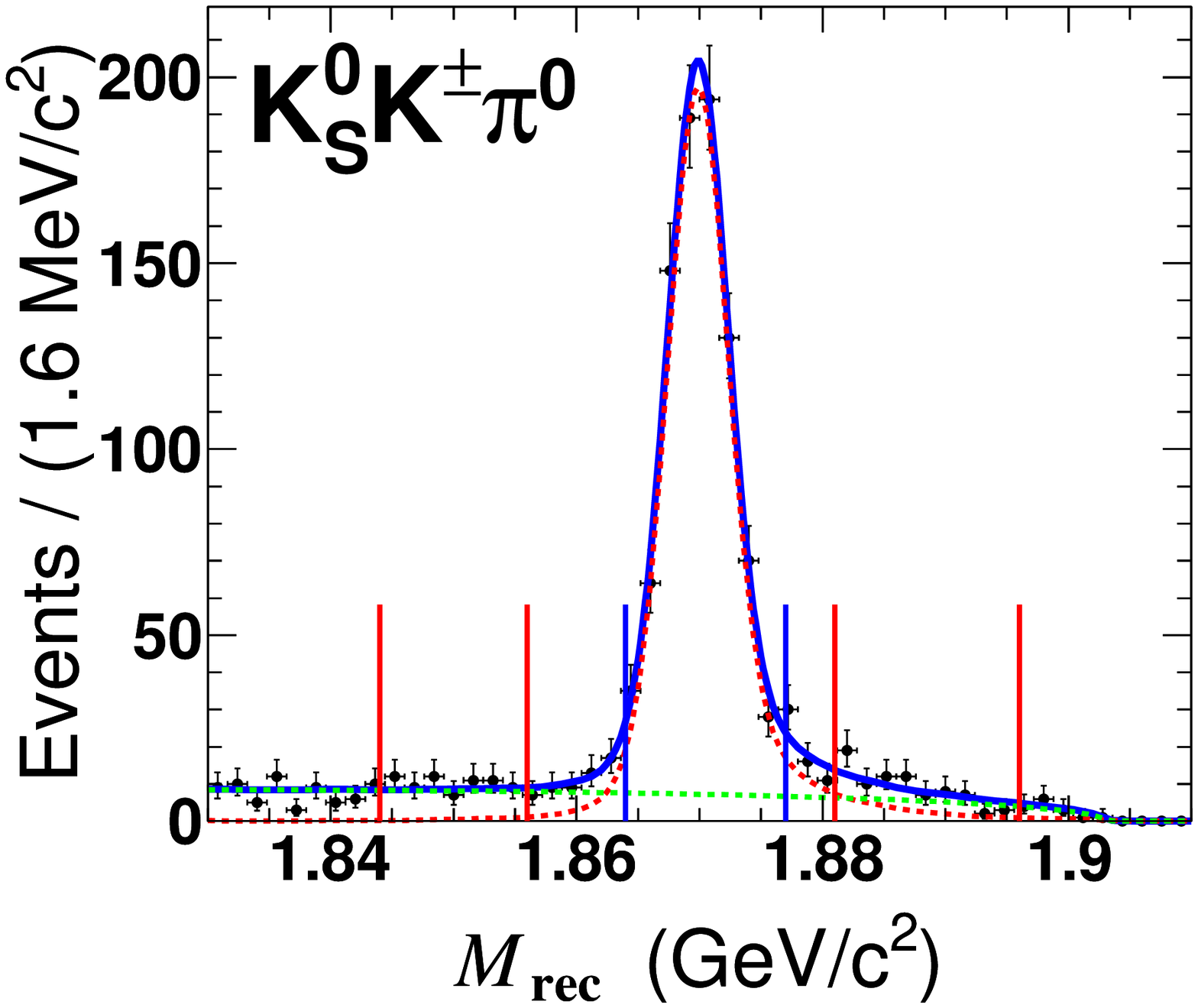}
  \includegraphics[width=0.49\linewidth]{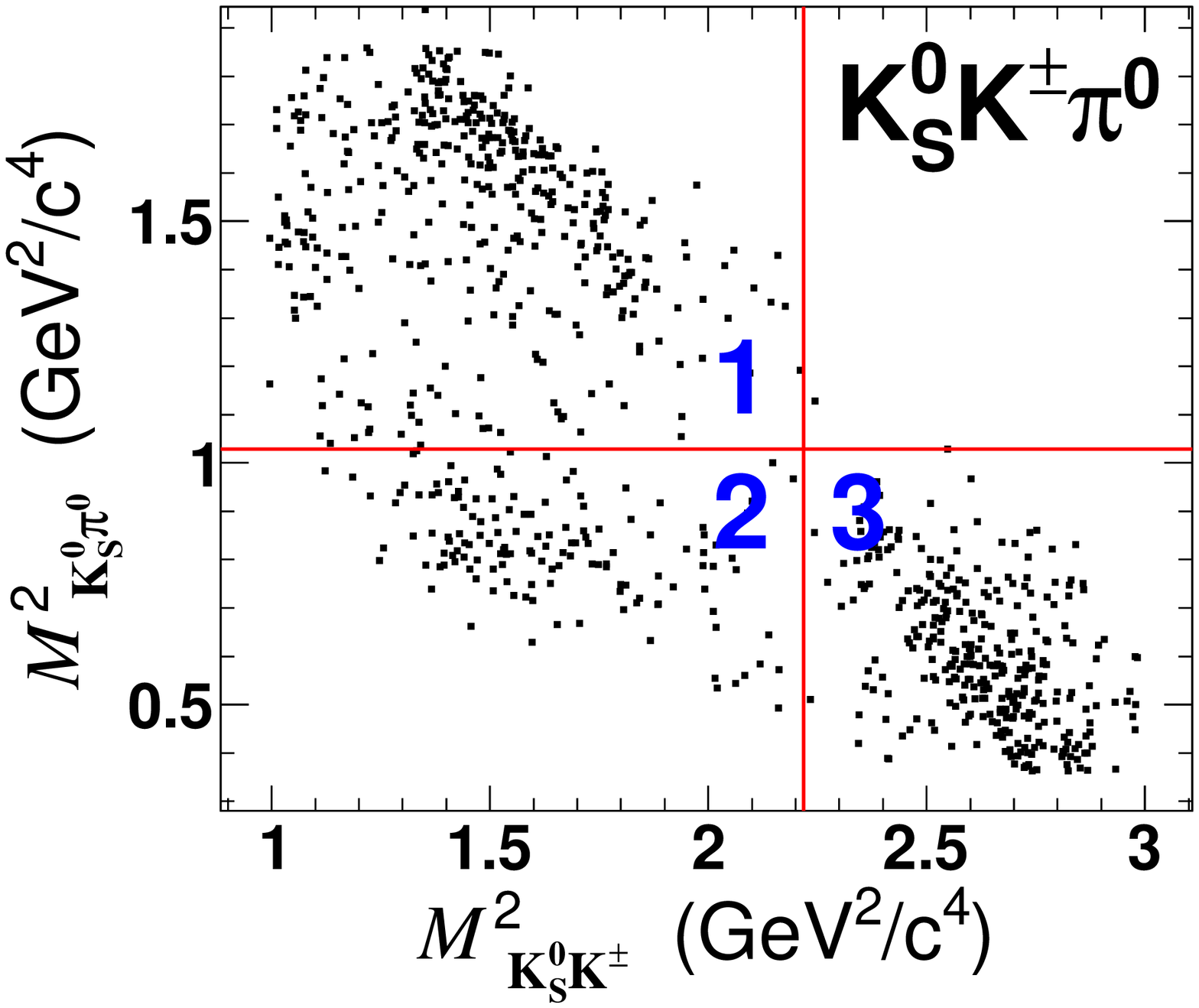}
  \includegraphics[width=0.49\linewidth]{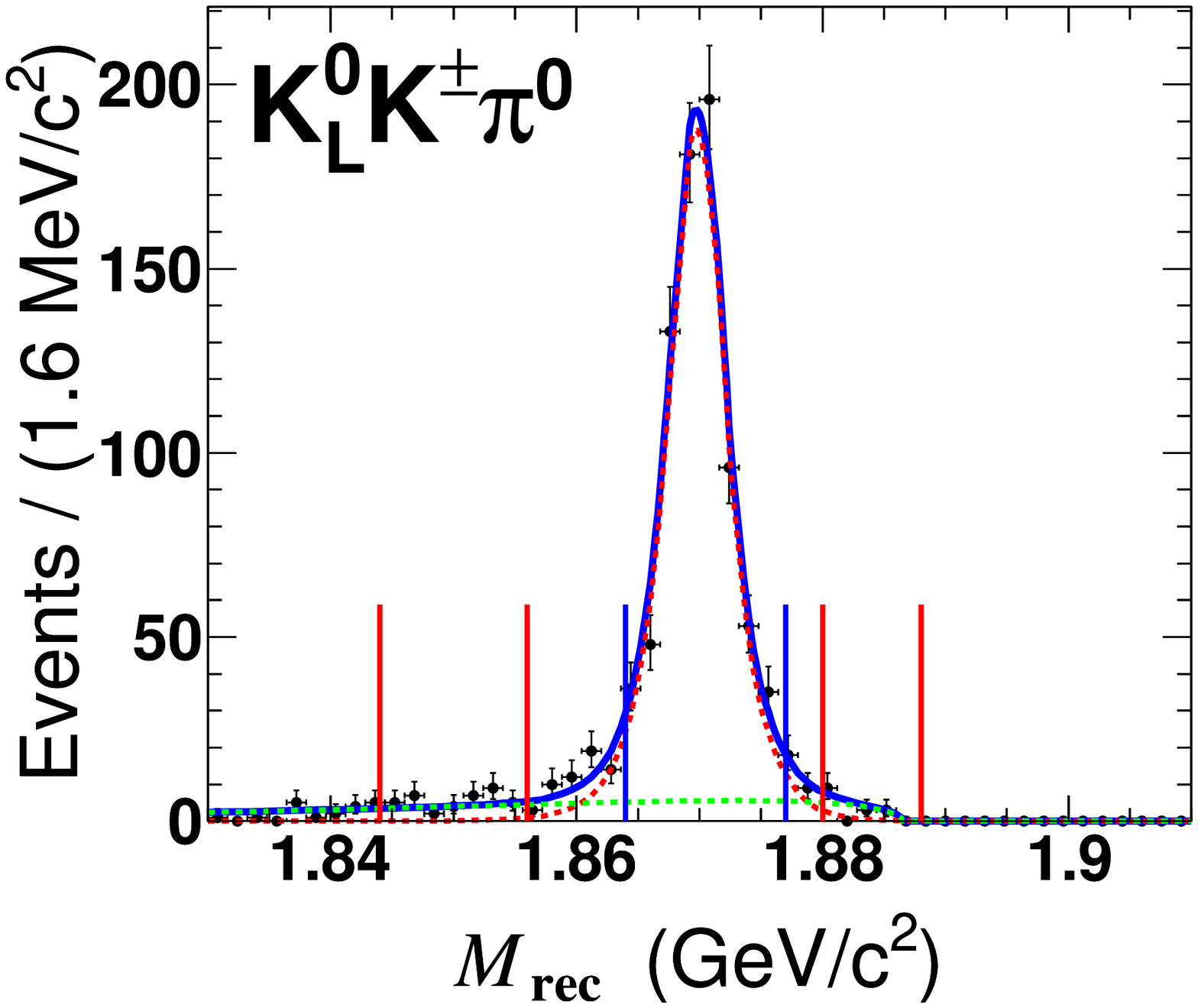}
  \includegraphics[width=0.49\linewidth]{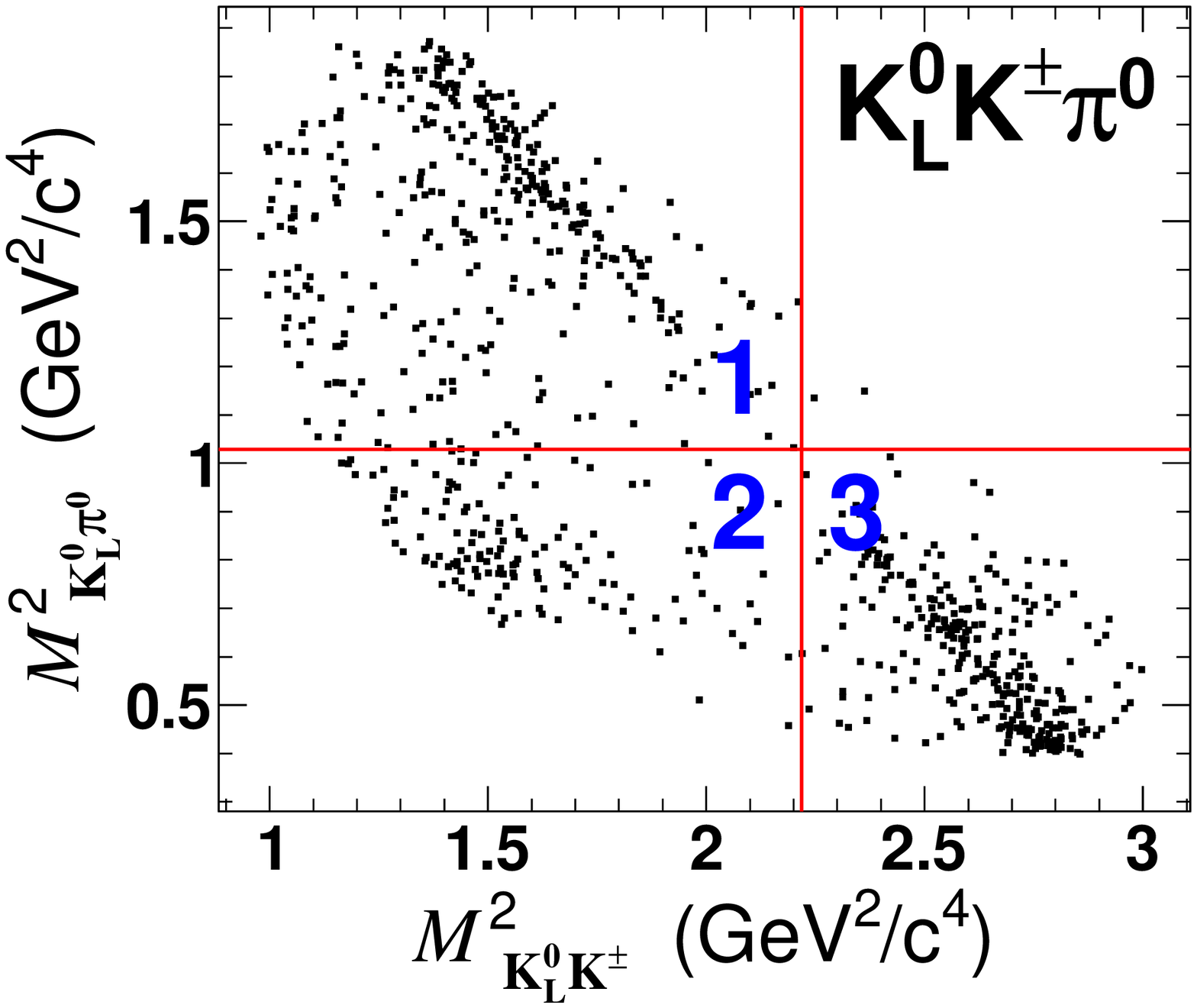}
  \caption{(Color online) (Left) Fits to the $M_{\rm rec}$ distributions of the $D^\pm\rightarrow K_{S,L}^0 K^\pm\pi^0$ candidate events, where the regions between the pairs of blue and red lines denote the signal and sideband regions, respectively. (Right) The Dalitz distribution of $M_{K_{S,L}^0 K^{\pm}}^2~(X)$ versus $M_{K_{S,L}^0 \pi^0}^2~(Y)$ for the events in the $M_{\rm rec}$ signal region. The Dalitz PHSP is divided into three regions, $i.e.$, region 1 with $X<2.22$~GeV$^2/c^4$ and $Y>1.03$~GeV$^2/c^4$, region 2 with $X < 2.22$~GeV$^2/c^4$ and $Y < 1.03$~GeV$^2/c^4$, region 3 with $X > 2.22$~GeV$^2/c^4$ and $Y < 1.03$~GeV$^2/c^4$.}
\label{MrecAndDp}
\end{figure}

Figure \ref{MrecAndDp} shows the fits to the $M_{\rm rec}$ distributions and the Dalitz plot of event candidates in the $M_{\rm rec}$ signal region defined as $(1.864,1.877)$~GeV/$c^2$. In the $M_{\rm rec}$ distribution of $D^\pm\rightarrow K_S^0 K^\pm\pi^0$, there is a significant tail above the $D^+$ mass due to ISR effects. For ISR events in $D^\pm\rightarrow K_L^0 K^\pm\pi^0$, the momentum of the $K_L^0$ becomes larger than what it should be due to the constraint of $\Delta E = 0$, which leads to a significant tail below the $D^\pm$ mass in the $M_{\rm rec}$ distribution. The $M_{\rm rec}$ distributions are fitted with a MC-derived signal shape convolved with a Gaussian function for the signal, together with an ARGUS function for the combinatorial background.

The Dalitz plot of $D^\pm\rightarrow K_{S,L}^0 K^\pm\pi^0$ is further divided into three regions to examine the $CP$ asymmetries. The DT yields in data are obtained by counting the numbers of events in the individual Dalitz plot regions in the $M_{\rm rec}$ signal region, and then subtract the numbers of background events in the $M_{\rm rec}$ sideband regions (shown in Fig.~\ref{MrecAndDp}). MC studies show that the peaking backgrounds in the study of $D^\pm\rightarrow K_{S}^0 K^\pm\pi^0$ are negligible. For the study of $D^\pm\rightarrow K_{L}^0 K^\pm\pi^0$, however, there are non-negligible peaking backgrounds dominated by $D^\pm\rightarrow K_S^0 K^\pm\pi^0$ with $K_S^0 \rightarrow \pi^0\pi^0$. These peaking backgrounds are estimated by MC simulations as described previously and are also subtracted from the data DT yields.

The background-subtracted DT yields in data $N_{\rm DT}$, the signal efficiencies $\epsilon$, the calculated branching fractions $\mathcal{B}$ and the $CP$ asymmetries $\mathcal{A}_{CP}$ in the different Dalitz plot regions are summarized in Tables \ref{DPcpresult1} and \ref{DPcpresult2}. Here, the branching fractions and the $CP$ asymmetries are calculated by Eq.~(\ref{BF}) and Eq.~(\ref{acp}), respectively. The corresponding systematic uncertainties are assigned after considering the different behaviors of $K^\pm$ and $K_{S,L}^0$ reconstruction in the detector. We use the same method as described in Sec.~\ref{sysBr} to estimate the systematic uncertainties on the $CP$ asymmetries in the individual Dalitz plot regions, all of which are listed in Table \ref{syscp2}. No evidence for $CP$ asymmetry is found in individual regions.

\begin{table}[hbtp]
  \centering
  \footnotesize
  \caption{Background-subtracted DT yields in data~$(N_{\rm DT})$ and detection efficiencies~$(\epsilon)$ in different Dalitz plot regions for $D^\pm\rightarrow K_{S,L}^0 K^\pm\pi^0$, where the uncertainties are statistical only.}
  \begin{spacing}{1.29}
  \begin{tabular}{L{0.1\linewidth}C{0.17\linewidth}C{0.2\linewidth}C{0.06\linewidth}C{0.17\linewidth}C{0.2\linewidth}}
 \hline\hline
 Region      &  $N_{\rm DT}$ & $\epsilon$ ($\%$)   &  &   $N_{\rm DT}$ & $\epsilon$ ($\%$)  \\
 \hline
             &  \multicolumn{2}{c}{$K_S^0K^+\pi^0$} & & \multicolumn{2}{c}{$K_S^0K^-\pi^0$}   \\

 1  &  201 $\pm$  15 &  9.25 $\pm$ 0.18 &   &   189   $\pm$     14 &   9.11 $\pm$ 0.18       \\
 2  &  50  $\pm$  8  &  13.80 $\pm$ 0.66 &  &   59    $\pm$     9  &   13.45 $\pm$ 0.66      \\
 3  &  164 $\pm$  14 &  11.68 $\pm$ 0.21 &  &   146   $\pm$     13 &   11.68 $\pm$ 0.21      \\

 \hline

 &  \multicolumn{2}{c}{$K_L^0K^+\pi^0$} & & \multicolumn{2}{c}{$K_L^0K^-\pi^0$}   \\

 1  &  177   $\pm$     14 &  8.04  $\pm$ 0.17 &  &   176    $\pm$    14 &   8.23  $\pm$ 0.17       \\
 2  &  51    $\pm$     8  &  13.29 $\pm$ 0.64 &  &   49     $\pm$    8  &   13.08 $\pm$ 0.64      \\
 3  &  146   $\pm$     13 &  10.13 $\pm$ 0.19 &  &   155    $\pm$    13 &   9.68 $\pm$ 0.19      \\

\hline
\hline
 \end{tabular}
 \end{spacing}
\label{DPcpresult1}
\end{table}

\begin{table}[hbtp]
  \centering
  \footnotesize
  \caption{Branching fractions~$(\mathcal{B})$ and $CP$ asymmetries~$(\mathcal{A}_{CP})$ in different Dalitz plot regions for $D^\pm\rightarrow K_{S,L}^0 K^\pm\pi^0$, where the first and second uncertainties are statistical and systematic, respectively.}
  \begin{spacing}{1.29}
  \begin{tabular}{L{0.08\linewidth}C{0.30\linewidth}C{0.30\linewidth}C{0.26\linewidth}}
 \hline\hline
   Region      &  $\mathcal{B}(D^+)$ ($\times10^{-3}$)      &    $\mathcal{B}(D^-)$ ($\times10^{-3}$)   &    $\mathcal{A}_{CP}$ ($\%$)  \\
  \hline
    &  $K_S^0 K^+ \pi^0$   &   $K_S^0 K^- \pi^0$   &    \\
 1  &  2.86 $\pm$  0.22 $\pm$ 0.10   &    2.75   $\pm$   0.21  $\pm$  0.09   &     2.0  $\pm$  5.4  $\pm$  2.4 \\
 2  &  0.48 $\pm$  0.08 $\pm$ 0.02   &    0.58   $\pm$   0.09  $\pm$  0.02   &     -9.4 $\pm$  11.3 $\pm$  2.7 \\
 3  &  1.85 $\pm$  0.16 $\pm$ 0.05   &    1.65   $\pm$   0.15  $\pm$  0.04   &     -5.7 $\pm$  6.3  $\pm$  1.8 \\
 \hline

    &  $K_L^0 K^+ \pi^0$   &    $K_L^0 K^- \pi^0$  &      \\

 1  &  2.89   $\pm$     0.24   $\pm$  0.08&    2.83    $\pm$    0.23  $\pm$   0.06&  1.0  $\pm$   5.8 $\pm$ 1.7 \\
 2  &  0.51   $\pm$     0.08   $\pm$  0.01&    0.50    $\pm$    0.08  $\pm$   0.01&  1.0  $\pm$   11.2 $\pm$ 1.4\\
 3  &  1.90   $\pm$     0.17   $\pm$  0.03&    2.12    $\pm$    0.18  $\pm$   0.03&  -5.5  $\pm$   6.1 $\pm$ 1.1\\
\hline
\hline
 \end{tabular}
 \end{spacing}
\label{DPcpresult2}
\end{table}

\begin{table}[hbtp]
  \centering
  \footnotesize
  \caption{Systematic uncertainties ($\%$) of the $CP$ asymmetries in different Dalitz plot regions for $D^\pm\rightarrow K_{S,L}^0 K^\pm\pi^0$.}\label{syscp2}
  \begin{spacing}{1.29}
   \begin{tabular}{L{0.3\linewidth}C{0.085\linewidth}C{0.085\linewidth}C{0.085\linewidth}  C{0.06\linewidth}C{0.085\linewidth}C{0.085\linewidth}C{0.085\linewidth}}
 \hline\hline
 Source &   1 & 2 &3     &   & 1 & 2 &3  \\
 \hline
 &  \multicolumn{3}{c}{$K_S^0K^+\pi^0$} & & \multicolumn{3}{c}{$K_S^0K^-\pi^0$}   \\
 $K^\pm$ tracking                &  2.5  &   1.4 & 1.1 & & 1.8 & 1.2 & 1.1        \\
 $K^{\pm}$ PID                   &  0.3  &   0.4 & 0.5 & & 0.6 & 0.3 & 0.2        \\
 $K_S^0$ reconstruction          &  2.6  &   3.5 & 2.3 & & 2.8 & 3.3 & 2.3        \\
 \quad Total                     &  3.6  &   3.8 & 2.6 & & 3.4 & 3.5 & 2.6        \\ \hline

 &  \multicolumn{3}{c}{$K_L^0K^+\pi^0$} & & \multicolumn{3}{c}{$K_L^0K^-\pi^0$}  \\
$K^\pm$ tracking               &  2.3 & 1.5 & 1.2 & & 1.7 & 1.4 & 1.1  \\
$K^\pm$ PID                    &  0.2 & 0.4 & 0.4 & & 0.6 & 0.1 & 0.1  \\
$K_L^0$ reconstruction         &  1.3 & 2.3 & 1.0 & & 1.3 & 2.2 & 1.0  \\
\quad Total                    &  2.6 & 2.8 & 1.6 & & 2.2 & 2.6 & 1.5  \\
\hline
\hline
 \end{tabular}
 \end{spacing}
\end{table}

\section{SUMMARY}
Using an $e^+ e^-$ collision data sample of 2.93 fb$^{-1}$ taken at $\sqrt{s}=3.773$~GeV with the BESIII detector, we present the measurements of the absolute branching fraction $\mathcal{B}(D^+\rightarrow K_S^0 K^+)$ = \brksk, which is in agreement with the CLEO result \cite{cleo-ksk}, and the three other absolute branching fractions $\mathcal{B}(D^+\rightarrow K_S^0 K^+\pi^0)$ = \brkskpi, $\mathcal{B}(D^+\rightarrow K_L^0 K^+)$ = \brklk,  $\mathcal{B}(D^+\rightarrow K_L^0 K^+\pi^0)$ = \brklkpi, which are measured for the first time. We also determine the direct $CP$ asymmetries for the four SCS decays and, for the decays $D^+ \rightarrow K_{S,L}^0 K^+\pi^0$,  also in different Dalitz plot regions. No evidence for direct $CP$ asymmetry is found. Theoretical calculations \cite{DtoPP} are in agreement with our measurements $\mathcal{B}(D^+\rightarrow K_{S,L}^0 K^+)$. Our measurements are helpful for the understanding of the SU(3)-flavor symmetry and its breaking mechanisms, as well as for $CP$ violation in hadronic $D$ decays \cite{cpscs1,SU3,NonLep,DtoPP}.

\section{ACKNOWLEDGMENTS}

The BESIII collaboration thanks the staff of BEPCII and the IHEP computing center for their strong support. This work is supported in part by National Key Basic Research Program of China under Contract No. 2015CB856700; National Natural Science Foundation of China (NSFC) under Contracts Nos. 11235011, 11322544, 11335008, 11425524, 11635010; the Chinese Academy of Sciences (CAS) Large-Scale Scientific Facility Program; the CAS Center for Excellence in Particle Physics (CCEPP); the Collaborative Innovation Center for Particles and Interactions (CICPI); Joint Large-Scale Scientific Facility Funds of the NSFC and CAS under Contracts Nos. U1232201, U1332201, U1532257, U1532258; CAS under Contracts Nos. KJCX2-YW-N29, KJCX2-YW-N45, QYZDJ-SSW-SLH003; 100 Talents Program of CAS; National 1000 Talents Program of China; INPAC and Shanghai Key Laboratory for Particle Physics and Cosmology; German Research Foundation DFG under Contracts Nos. Collaborative Research Center CRC 1044, FOR 2359; Istituto Nazionale di Fisica Nucleare, Italy; Koninklijke Nederlandse Akademie van Wetenschappen (KNAW) under Contract No. 530-4CDP03; Ministry of Development of Turkey under Contract No. DPT2006K-120470; National Science and Technology fund; The Swedish Research Council; U. S. Department of Energy under Contracts Nos. DE-FG02-05ER41374, DE-SC-0010118, DE-SC-0010504, DE-SC-0012069; University of Groningen (RuG) and the Helmholtzzentrum fuer Schwerionenforschung GmbH (GSI), Darmstadt; WCU Program of National Research Foundation of Korea under Contract No. R32-2008-000-10155-0. This paper is also supported by the NSFC under Contract Nos. 11475107.

\end{spacing}


\begin{thebibliography}{**}


\bibitem{cpscs1} K. Waikwok and S. Rosen, \href{http://www.sciencedirect.com/science/article/pii/037026939391843C}{Phys. Lett. B {\bf 298}, 413 (1993)}.

 \bibitem{SU3} Y. Grossman and D. J. Robinson, \href{https://link.springer.com/article/10.1007%2FJHEP01%282011%29132}{J. High Energy Phys. {\bf 1101}, 132 (2011)}.


\bibitem{NonLep}  F. -S. Yu, X. -X. Wang, and C. -D. L$\ddot{\rm u}$, \href{https://journals.aps.org/prd/abstract/10.1103/PhysRevD.84.074019}{Phys. Rev. D {\bf 84}, 074019 (2011)}.

\bibitem{DtoPP} H.-n. Li, C.-D. Lu, and F.-S. Yu, \href{https://journals.aps.org/prd/abstract/10.1103/PhysRevD.86.036012}{Phys. Rev. D {\bf 86},
036012 (2012)}.

\bibitem{DT1} R. M. Baltrusaitis {\it et al.} (MARK-III Collaboration), \href{http://journals.aps.org/prl/abstract/10.1103/PhysRevLett.56.2140}{Phys. Rev. Lett. {\bf 56}, 2140 (1986)}.

\bibitem{DT2} J. Adler {\it et al.} (MARK-III Collaboration), \href{http://journals.aps.org/prl/abstract/10.1103/PhysRevLett.60.89}{Phys. Rev. Lett. {\bf 60}, 89 (1988)}.

\bibitem{lum} M. Ablikim {\it et al.} (BESIII Collaboration), \href{http://iopscience.iop.org/article/10.1088/1674-1137/37/12/123001/pdf}{Chin. Phys. C {\bf 37}, 123001 (2013)}; \;  \href{http://www.sciencedirect.com/science/article/pii/S0370269315008990}{Phys. Lett. B $\bm{753}$, 629 (2016)}.

\bibitem{detector} M. Ablikim {\it et al.} (BESIII Collaboration), \href{http://www.sciencedirect.com/science/article/pii/S0168900209023870} {Nucl. Instrum. Methods Phys. Res., Sect. A {\bf 614}, 345 (2010)}.

\bibitem{BEPCII} C. Zhang,  \href{http://link.springer.com/article/10.1007%2Fs11433-010-4139-2}{Sci. China. Phys., Mech. Astron., $\bm{53}$, 2084 (2010)}.

\bibitem{geant4} S. Agostinelli {\it et al.} (GEANT4 Collaboration),  \href{http://www.sciencedirect.com/science/article/pii/S0168900203013688}{Nucl. Instrum. Methods Phys. Res., Sect. A {\bf 506}, 250 (2003)}.

\bibitem{KKMC} S. Jadach, B. F. L. Ward, and Z. Was,  \href{http://journals.aps.org/prd/abstract/10.1103/PhysRevD.63.113009}{Phys. Rev. D {\bf 63}, 113009 (2001)}.

\bibitem{FSR} E. Barberio and Z. Was, \href{http://www.sciencedirect.com/science/article/pii/0010465594900744}{Comput. Phys. Commun. {\bf 79}, 291 (1994)}.

\bibitem{EVENT1}D. J. Lange, \href{http://www.sciencedirect.com/science/article/pii/S0168900201000894}{Nucl. Instrum. Methods Phys. Res., Sect. A {\bf 462}, 152 (2001)}.

\bibitem{EVENT2}R. G. Ping, \href{http://dx.doi.org/10.1088/1674-1137/32/8/001}{Chin. Phys. C {\bf 32}, 599 (2008)}.

\bibitem{PDG} M. Tanabashi {\it et al.} (Particle Data Group) \href{http://iopscience.iop.org/article/10.1088/1674-1137/38/9/090001/meta}{Phys. Rev. D. {\bf 98}, 030001 (2018)}.

\bibitem{kkpi} P. Rubin {\em et al.} (CLEO Collaboration), \href{http://journals.aps.org/prd/abstract/10.1103/PhysRevD.78.072003}{Phys. Rev. D. {\bf 78}, 072003 (2008)}.

\bibitem{klenu} M. Ablikim {\it et al.} (BESIII Collaboration), \href{http://journals.aps.org/prd/abstract/10.1103/PhysRevD.92.112008}{Phys. Rev. D. {\bf 92}, 112008 (2015)}.

\bibitem{ARGUS} H. Albrecht {\it et al.} (ARGUS Collaboration), \href{http://www.sciencedirect.com/science/article/pii/037026939091293K}{Phys. Lett. B {\bf 241}, 278 (1990)}.

\bibitem{cleo-ksk} G. Bonvicini {\em et al.} (CLEO Collaboration), \href{http://journals.aps.org/prd/abstract/10.1103/PhysRevD.77.091106}{Phys. Rev. D. {\bf 77}, 091106(R) (2008)}.



\end{thebibliography}
\end{document}